\documentclass[aps,prc,superscriptaddress,reprint]{revtex4-1}
\usepackage{graphicx} 
\usepackage{epstopdf}
\usepackage{longtable,booktabs}
\usepackage{dcolumn} 
\usepackage{bm}
\usepackage{hyperref}
\hypersetup{
    colorlinks=true,
    linkcolor=blue,
    filecolor=gray,
    urlcolor=blue,
    citecolor=blue,
}
\begin{document}
\title{Production mechanism of neutron-deficient actinide isotopes in complete fusion reactions and multinucleon transfer reactions}

\author{Peng-Hui Chen}
\affiliation{South China University of Technology, Guangzhou 510640, China}
\author{Fei Niu}
\affiliation{South China University of Technology, Guangzhou 510640, China}
\author{Zhao-Qing Feng}
\email{Corresponding author: fengzhq@scut.edu.cn}
\affiliation{South China University of Technology, Guangzhou 510640, China}
\date{\today}
\begin{abstract}
Within the dinuclear system model, unknown neutron-deficient isotopes Np, Pu, Am, Cm, Bk, Cf, Es, Fm are investigated in $^{40}$Ca, $^{36,40}$Ar, $^{32}$S, $^{28}$Si,$^{24}$Mg induced fusion-evaporation reactions and multinucleon transfer reactions with radioactive beams $^{59}$Cu,$^{69}$As,$^{90}$Nb,$^{91}$Tc, $^{94}$Rh, $^{105,110}$Sn, $^{118}$Xe induced with $^{238}$U near Coulomb barrier energies. The production cross sections of compound nuclei in the fusion-evaporation reactions and fragments yields in the multinucleon transfer reactions are calculated within the model. A statistical approach is used to evaluate the survival probability of excited nuclei via the both reaction mechanisms. A dynamical deformation is implemented into the model in the dissipation process. It is found that charge particle channels (alpha and proton) dominate in the decay process of proton-rich nuclides and the fusion-evaporation reactions are favorable to produce the new neutron-deficient actinide isotopes. The total kinetic energies and angular spectra of primary fragments are strongly dependent on colliding orientations.

\begin{description}
\item[PACS number(s)]
25.70.Jj, 24.10.-i, 25.60.Pj
\end{description}
\end{abstract}

\maketitle

\section{Introduction}

In the past decades, studies on producing neutron-deficient actinide nuclei have performed continually in Institute of Modern Physics (IMP, Lanzhou), Flerov Laboratory of Nuclear Reactions (FLNR, Dubna) and Lawrence Berkeley National Laboratory (LBNL, Berkeley). The new proton-rich isotopes were synthesized in experiments through fusion-evaporation reactions by detecting the alpha decay chains \cite{yang18,yang15,ma15,lau99}. The products in the fusion-evaporation (FE) reactions are closely associated with the projectile-target mass asymmetry. The combination of light projectile with a heavy target is used for for creating the proton-rich actinide isotopes through fusion-evaporation reaction, for instance, $^{36}$Ar + $^{208}$Pb $\rightarrow$ $^{244}$Fm, $^{40}$Ca + $^{196}$Hg $\rightarrow$ $^{236}$Fm etc. On the other hand, the multinucleon transfer (MNT) reactions based on neutron-deficient beams might be a possible way. In the experiment for synthesizing superheavy nuclei with $^{48}$Ca + $^{248}$Cm \cite{Deva15}, five new neutron-deficient isotopes $^{216}$U, $^{219}$Np, $^{223}$Am, $^{229}$Am, $^{233}$Bk have been identified, in which the MNT process proceeds in the reactions. It has advantage that the products are formed with the wide mass regime owing to broad excitation function in the MNT products. The neutron-deficient radioactive beams such as $^{105, 110, 115}$Sn  $^{58}$Cu, $^{69}$As, $^{90}$Nb, $^{94}$Rh, $^{118}$Xe can be generated in radioactive beams facilities, for instance, Beijing Rare Ion beam Facility (BRIF), Beijing Isotope-Separation On Line (BISOL) and Radioactive Ion Beam Facility (BIBM, RIKEN). The MNT reaction within neutron-deficient beams might be favorable to approach neutron-deficient actinide region due to isospin relaxation. On the other hand, the properties of neutron-deficient heavy isotopes is crucial in exploring proton drip line and shell evolution. The MNT reactions might be a possible way to produce neutron-deficient heavy isotopes in the nuclide chart, instead of fusion-evaporation reactions.

Following the motivation for producing heavy new isotopes, several models have been developed for describing the transfer reactions, i.e., the dynamical model based on multidimensional Langevin equations \cite{Za07,Za08}, the time-dependent Hartree-Fock (TDHF) approach \cite{Gola09,Seki16,Gu18,Ji18}, the GRAZING model \cite{Wint94,bib:6}, the improved quantum molecular dynamics (ImQMD) model \cite{Zha15}, and the dinuclear system (DNS) model\cite{Fe09,bib:8}, etc. Some interesting issues have been stressed, e.g., the production cross sections of new isotopes, total kinetic energy spectra of transfer fragments, structure effect on the fragment formation, primary products angle distributions. There are still some open problems for the transfer reactions, i.e., including the mechanism of preequilibrium cluster emission, the stiffness of nuclear surface during the nucleon transfer process, the mass limit of new isotopes with stable heavy target nuclides, etc. Traditionally, neutron-deficient heavy nuclei have been produced through fusion-evaporation mechanism, which has a shortcoming for producing extreme neutron-deficient actinide nuclei due to small fusion probability. The MNT reactions might provide a possible way to approach neutron-deficient actinide isotopes close to proton drip line.

The transfer reactions and deep inelastic heavy-ion collisions were extensively investigated in experiments since 1970s, in which the new neutron-rich isotopes of light nuclei and pronton-rich actinide nuclei were observed \cite{Art71, Art73, Art74, Hil77, Gla79, Moo86, Wel87}. The reaction mechanism and fragment formation were investigated thoroughly, i.e., the energy and angular momentum dissipation, two-body kinematics, shell effect, fission of actinide nuclei etc. Recently, more measurements have been performed at different laboratories for creating the neutron-rich heavy nuclei, e.g., the reactions of $^{136}$Xe+$^{208}$Pb \cite{Ko12,Ba15}, $^{136}$Xe+$^{198}$Pt \cite{Wa15}, $^{156,160}$Gd+$^{186}$W \cite{Ko17}, $^{238}$U+$^{232}$Th \cite{Wu18}. The MNT reactions with radioactive beams are feasible for producing new isotopes owing to the isospin equilibrium \cite{Das94,Lov19}.

In this work, the $^{40}$Ca, $^{36}$Ar, $^{32}$S, $^{28}$Si, $^{24}$Mg induced fusion-evaporation reactions and the MNT reactions with the combinations of $^{105, 110,115,120,125,130}$Sn  $^{58}$Cu, $^{69}$As, $^{90}$Nb, $^{94}$Rh, $^{118}$Xe with $^{238}$U are calculated with the DNS model. The article is organized as follows: In Sec. \ref{sec2} we give a brief description of the DNS model. Calculated results and discussion are presented in Sec. \ref{sec3}. Summary is concluded in Sec. \ref{sec4}.

\section{Model description}\label{sec2}

The DNS concept was proposed by Volkov for describing the deep inelastic heavy-ion collisions \cite{Vo78}, in which a few nucleon transfer was treated. Application of the approach to superheavy nucleus formation via massive fusion reactions in competition with the quasifission process was used for the first time by Adamian \emph{et al.} \cite{Ad97,Ad98}. The modifications of the relative motion energy and angular momentum of two colliding nuclei coupling to nucleon transfer within the DNS concept were performed by the Lanzhou Group \cite{Li03,Fe05,Fe06}. The production cross sections of SHN, quasi-fission and fusion-fission dynamics have been extensively investigated within the DNS model \cite{Fe09b,Fe10}. The dynamical evolution of colliding system sequentially proceeds the capture process by overcoming the Coulomb barrier to form the DNS, relaxation process of the relative motion energy, angular momentum, mass and charge asymmetry etc within the potential energy surface and the de-excitation of primary fragments.

The distribution probability is obtained by solving a set of master equations numerically in the potential energy surface of the DNS. The time evolution of the distribution probability $P(Z_{1},N_{1},E_{1},\beta, t)$for fragment 1 with proton number $Z_{1}$, neutron number $N_{1}$, excitation energy $E_{1}$, quadrupole deformation $\beta$ is described by the following master equations:
\begin{eqnarray}
&&\frac{d P(Z_1,N_1,E_1,\beta,t)}{d t} =  \sum \limits_{Z^{'}_1}  W_{Z_1,N_1,\beta;Z'_1,N_1,\beta}(t)  \nonumber \\
&& [d_{Z_1,N_1} P(Z'_1,N_1,E'_1,\beta,t) - d_{Z'_1,N_1}P(Z_1,N_1,E_1,\beta,t)] +  \nonumber  \\
&& \sum \limits_{N'_1}  W_{Z_1,N_1,\beta;Z_1,N'_1,\beta}(t) \nonumber   \\
&& [d_{Z_1,N_1}P(Z_1,N'_1,E'_1,\beta,t) - d_{Z_1,N'_1}P(Z_1,N_1,E_1,\beta,t)]
\end{eqnarray} 
Here the $W_{Z_{1},N_{1},\beta;Z^{'}_{1},N_{1},\beta}$($W_{Z_{1},N_{1},\beta,;Z_{1},N^{'}_{1},\beta}$) is the mean transition probability from the channel($Z_{1},N_{1},E_{1},\beta$) to ($Z^{'}_{1},N_{1},E^{'}_{1},\beta$), [or ($Z_{1},N_{1},E_{1},\beta$) to ($Z_{1},N^{'}_{1},E^{'}_{1},\beta$)], and $d_{Z_{1},Z_{1}}$ denotes the microscopic dimension corresponding to the macroscopic state ($Z_{1},N_{1},E_{1}$).The sum is taken over all possible proton and neutron numbers that fragment $Z^{'}_{1}$,$N^{'}_{1}$ may take, but only one nucleon transfer is considered in the model with the relations $Z^{'}_{1}=Z_{1}\pm1$ and $N^{'}_{1}=N_{1}\pm1$. It is noticed that the decay of DNS is not taken into account because of vanishing the quasifission barrier, which was included in the fusion-evaporation reactions. Actually, the decay of the DNS has been effectively considered with shortening the interaction time.

The motion of nucleons in the interacting potential is governed by the single-particle Hamiltonian. The excited DNS opens a valence space in which the valence nucleons have a symmetrical distribution around the Fermi surface. Only the particles at the states within the valence space are actively for nucleon transfer. The transition probability is related to the local excitation energy and nucleon transfer, which is microscopically derived from the interaction potential in valence space as
\begin{eqnarray}
W_{Z_{1},N_{1};Z_{1}^{\prime},N_{1}}=&& \frac{\tau_{mem}(Z_{1},N_{1},E_{1};Z_{1}^{\prime},N_{1},E_{1}^{\prime})}{d_{Z_{1},N_{1}} d_{Z_{1}^{\prime},N_{1}}\hbar^{2}}  \times   \nonumber \\
&&\sum_{ii^{\prime}}|\langle  Z_{1}^{\prime},N_{1},E_{1}^{\prime},i^{\prime}|V|Z_{1},N_{1},E_{1},i \rangle|^{2}.
\end{eqnarray}
The transition coefficients determine the distribution width of the isotopic yields in the MNT reactions. The memory time $\tau_{mem}$ and interaction element $V$ can be seen in Ref. \cite{Ch17}. The similar approach is used for the neutron transition coefficient.

The averages on these quantities are performed in the valence space as follows \cite{No75}.
\begin{eqnarray}
\Delta \varepsilon_K = \sqrt{\frac{4\varepsilon^*_K}{g_K}},\quad
\varepsilon^*_K =\varepsilon^*\frac{A_K}{A}, \quad
g_K = A_K /12,
\end{eqnarray}
where the $\varepsilon^*$ is the local excitation energy of the DNS. The microscopic dimension for the fragment ($Z_{K},N_{K}$) is evaluated by the valence states $N_K$ = $g_K\Delta\varepsilon_K$ and the valence nucleons $m_K$ = $N_K/2$ ($K=1,2$) as
\begin{eqnarray}
 d(m_1, m_2) = {N_1 \choose m_1} {N_2 \choose m_2}.
\end{eqnarray}

In the relaxation process of the relative motion, the DNS will be excited by the dissipation of the relative kinetic energy. The local excitation energy is determined by the dissipation energy from the relative motion and the potential energy surface of the DNS as
\begin{eqnarray}
\varepsilon^{\ast}(t)=E^{diss}(t)-\left(U(\{\alpha\})-U(\{\alpha_{EN}\})\right).
\end{eqnarray}
The entrance channel quantities $\{\alpha_{EN}\}$ include the proton and neutron numbers, quadrupole deformation parameters and orientation angles being $Z_{P}$, $N_{P}$, $Z_{T}$, $N_{T}$, $R$, $\beta_{P}$, $\beta_{T}$, $\theta_{P}$, $\theta_{T}$ for projectile and target nuclei with the symbols of $P$ and $T$, respectively. The excitation energy $E_{1}$ for fragment (Z$_{1}$,N$_{1}$) is evaluated by $E_{1}=\varepsilon^{\ast}(t=\tau_{int})A_{1}/A$.

\begin{figure}
\includegraphics[width=1.\linewidth]{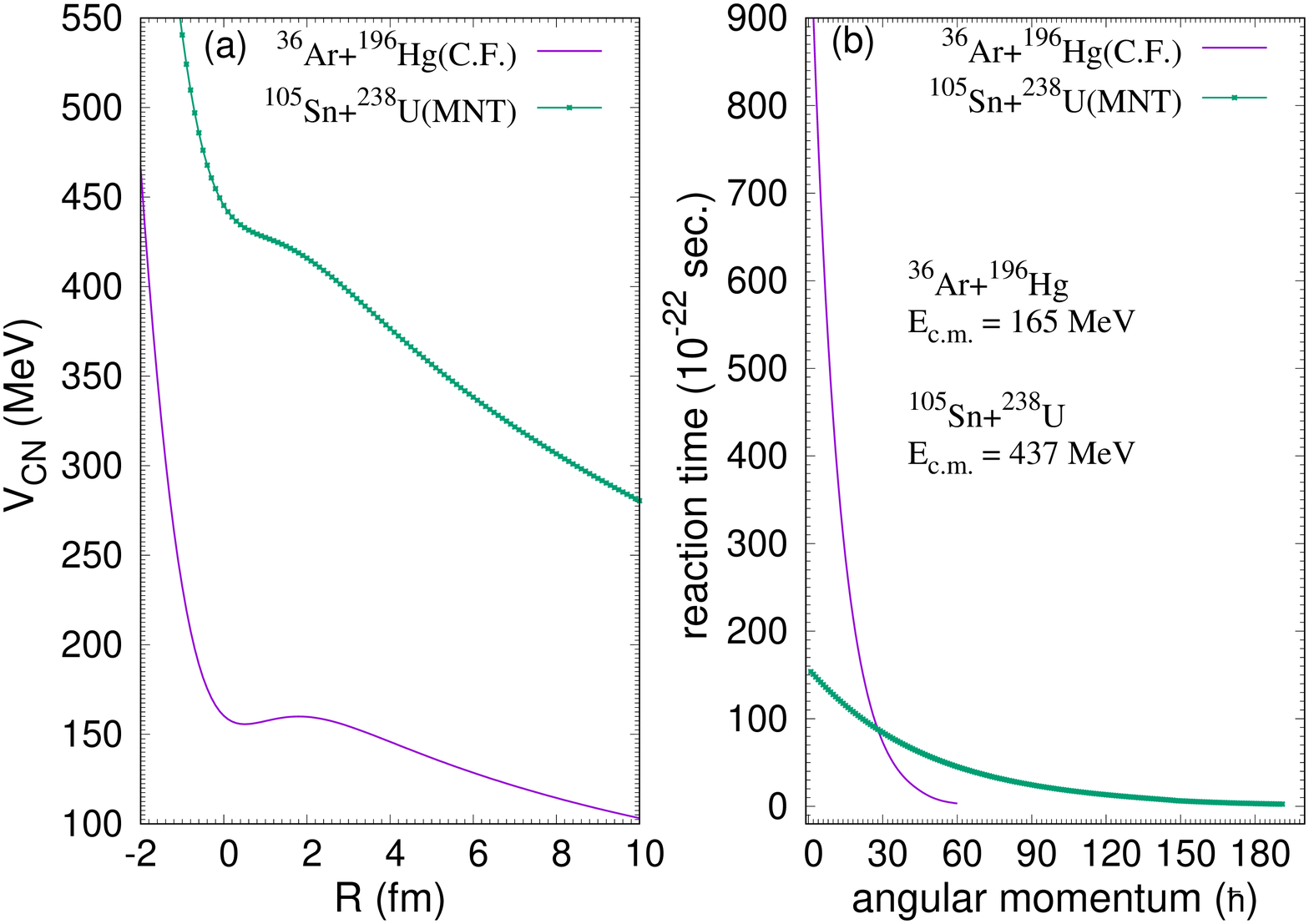}
\caption{\label{fig1}  (a) The interaction potentials and (b) angular momentum dependence of the reaction time in the fusion-evaporation reaction $^{36}$Ar+$^{196}$Hg and in the MNT reaction $^{105}$Sn+$^{238}$U.}
\end{figure}

The interaction time $\tau_{int}$ is obtained from the deflection function method \cite{Wo78}. The interaction potential is composed of Coulomb and nuclear potential which are calculated by Wong formula and double folding formalism \cite{Ch17}. The interaction potential energy distribution and interaction time for the systems of $^{36}$Ar + $^{196}$Hg (magenta line) and $^{105}$Sn+$^{238}$U (olive lines points) reactions are shown in Fig. \ref{fig1}. It should be noticed that there is no potential pocket for the heavy systems. The interaction decreases exponentially with increasing angular momentum. The existence of the pocket in the entrance channel is crucial for the compound nucleus formation in fusion reactions \cite{De02}. The barrier is taken as the potential value at the touching configuration and the nucleus-nucleus potential is calculated with the same approach in fusion reactions \cite{Fe06}. According to Fig. \ref{fig1}, we found that light systems have a longer interaction time due to potential pocket (Coulomb barrier), in comparison with heavy systems. The lifetime of the DNS is strongly reduced in the MNT reactions in comparison to the fusion-evaporation reactions, i.e., the half width value  of relaxation time being $50 \times 10^{-22}$s for the system $^{105}$Sn+$^{238}$U and $300 \times 10^{-22}$s for the reaction $^{36}$Ar+$^{196}$Hg.

The energy dissipated into the DNS is expressed as
\begin{eqnarray}
\label{eq4}
E^{diss}(t) = E_{c.m.} - B - \frac{<J(t)>[<J(t)>+1]\hbar ^2}{2\zeta}     \nonumber
\\ - <E_{rad}(J,t)>
\end{eqnarray}
Here the $E_{c.m.}$ and B are the centre of mass energy and Coulomb barrier, respectively. The radial energy is evaluated from
\begin{equation}
\label{eq41}
 <E_{rad}(J,t)> = E_{rad}(J,0) \exp{(-t/ \tau _r)}.
\end{equation}
The relaxation time of the radial motion $\tau _r$= 5 $\times 10^{-22}$ s and the radial energy at the initial state $E_{rad}(J,0) = E_{c.m.} - B - J_i(J_i +1) \hbar ^2 / (2 \zeta _{rel})$. The dissipation of the relative angular momentum is described by
\begin{equation}
\label{eq5}
<J(t)> = J_{st} +(J_i - J_{st}) \exp(-t/ \tau _J).
\end{equation}
The angular momentum at the sticking limit $J_{st} = J_i \zeta _{rel}/ \zeta _{tot}$ and the relaxation time $\tau _J = 15 \times 10^{-22}$ s. The $\zeta _{rel}$ and $\zeta _{tot}$ are the relative and total moments of inertia of the DNS, respectively, in which the quadrupole deformations are implemented \cite{Fe07b}. The initial angular momentum is set to be $J_i = J$ in the following work. In the relaxation process of the relative motion, the DNS will be excited by the dissipation of the relative kinetic energy.

\begin{figure}
\includegraphics[width=1.\linewidth]{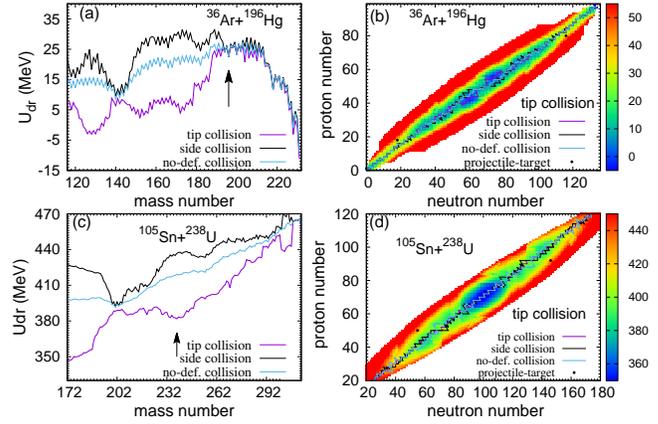}
\caption{\label{fig2} Potential surface energy of $^{36}$Ar+$^{196}$Hg and $^{105}$Sn+$^{238}$U with the tip-tip, side-side and no-deformation collisions. The entrance channels are marked by arrows and black solid circles.}
\end{figure}

The local excitation energy is determined by the excitation energy of the composite system and the potential energy surface (PES) of the DNS. The PES is evaluated by
\begin{eqnarray}\label{dri}
U_{dr}(t) = Q_{gg}+V_C(Z_1,N_1,\beta_1,Z_2,N_2,\beta_2,t)     \nonumber   \\
+ V_N(Z_1,N_1,\beta_1,Z_2,N_2,\beta_2,t) + V_{def}(t)
\end{eqnarray}
with
\begin{eqnarray}
V_{def}(t) = \frac{1}{2} C_1 (\beta_1 - \beta' _T (t) )^2 + \frac{1}{2} C_2 (\beta_2 - \beta' _P (t) )^2      \\
C_i = (\lambda-1) { (\lambda+2) R^2_N \delta - \frac{3}{2\pi}} \frac{Z^2e^2}{R_N(2\lambda+1)}
\end{eqnarray}
which satisfies the relation of $ Z_{1}+Z_{2}=Z $ and  $ N_{1}+N_{2}=N$ with the $Z$ and $N$ being the proton and neutron numbers of composite system, respectively. The symbol ${\alpha}$ denotes the quantities of $Z_{1}$, $N_{1}$, $Z_{2}$, $N_{2}$, $J$, $R$, $\beta_{1}$, $\beta_{2}$, $\theta_{1}$, $\theta_{2}$. The $B(Z_{i},N_{i}) (i=1,2)$ and $B(Z,N)$ are the negative binding energies of the fragment $(Z_{i},N_{i})$ and the composite system $(Z,N)$, respectively. The $\theta_{i}$ denotes the angles between the collision orientations and the symmetry axes of the deformed nuclei. Shown in Fig. \ref{fig2} is the PES in the tip-tip collisions of $^{105}$Sn+$^{238}$U and $^{36}$Ar+$^{196}$Hg. The DNS fragments towards the mass symmetric valley release the positive energy, which is available for nucleon transfer. The spectra exhibits a symmetric distribution for each isotopic chain. The valley in the PES is close to the $\beta$-stability line and enables the diffusion of the fragment probability. The entrance position of projectile and target nuclei is indicated by black dots in the PES contour graphs. The occupation probability diffuses from the entrance position to possible states once overcoming the local poential energy. The evolutions of quadrupole deformations of projectile-like and target-like fragments undergo from the initial configuration as
\begin{eqnarray}
\beta' _P(t) = \beta _ P \exp{-t/\tau_{\beta}} + \beta_1 [ 1 - \exp(-t/\tau_{\beta})],      \nonumber \\
\beta' _T (t) = \beta _ T \exp{-t/\tau_{\beta}} + \beta_2 [ 1 - \exp(-t/\tau_{\beta})]
\end{eqnarray}
with the deformation relaxation of $\tau_{\beta}=40\times10^{-22} \ s$.

\begin{figure}
\includegraphics[width=1.\linewidth]{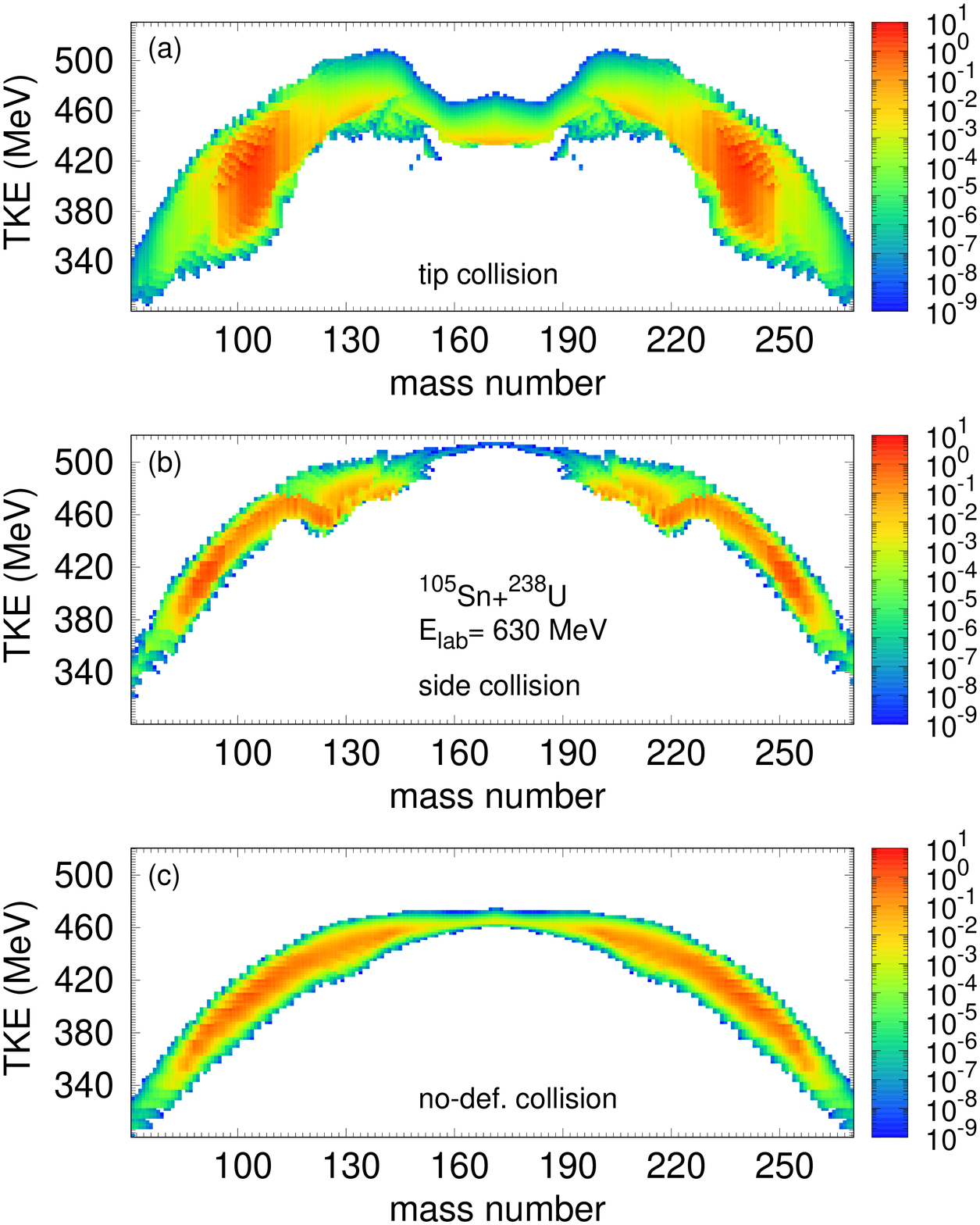}
\caption{\label{fig3} The total kinetic energy and mass distributions of the primary fragments produced in the MNT reactions of $^{105}$Sn+$^{238}$U collisions at E$_{lab}$ = 6 MeV/nucleon with the (a) tip-tip, (b) side-side and (c) no-deformation collisions, respectively.}
\end{figure}

The total kinetic energy (TKE) of the primary fragment is evaluated by
\begin{equation}
\label{eq7}
TKE(A_{1}) = E_{c.m.} + Q_{gg}(A_{1}) - E^{diss}(t=\tau_{int}),
\end{equation}
where $Q_{gg} = M_P + M_T - M_{PLF} -M_{TLF}$ and $E_{c.m.}$ being the incident energy in the center of mass frame. The mass $M_P$, $M_T$, $M_{PLF}$ and $M_{TLF}$ correspond to projectile, target, projectile-like fragment and target-like fragment, respectively. Figure \ref{fig3} shows the calculated total kinetic energy (TKE) and the mass distributions of the primary products with inclusive mass distribution for the $^{105}$Sn+ $^{238}$U reaction with three types of collision orientations at near-barrier energy of $E_{lab}$ = 6 MeV/nucleon. The TKE is highly dependent on the initial orientation of the deformed $^{105}$Sn and $^{238}$U nuclei, caused by PES.  The formation of DNS fragments tends to the symmetric pathway (quasifission process). The spectra exhibit a symmetric mass distribution because of the structure in the PES. We found that TKE and mass distribution with the tip-tip collision is wider than that in side-side and no-deformation collisions. The tail of TKE distribution can reach very low kinetic energy with small yields due to massive kinetic energy dissipation. The large yields of the fragments in the region from target position to doubly magic nucleus $^{208}$Pb is the most pronounced feature of the TKE distribution.

The cross sections of the survival fragments produced in the MNT reactions and the evaporation residue cross sections are evaluated by
\begin{eqnarray}
&&\sigma_{MNT}(Z_{1},N_{1},E_{c.m.})= \frac{\pi \hbar^{2}}{2\mu  E_{c.m.}}\sum_{J=0}^{J_{\max}} (2J+1)     \nonumber \\
&& \times  \int  f(B) T(E_{c.m.},J,B) \sum_{s}P(Z^{'}_{1},N^{'}_{1},E^{'}_{1},J^{'}_{1},B)   \nonumber \\
&& \times  W_{sur}(Z^{'}_{1},N^{'}_{1},E^{'}_{1},J^{'}_{1},s)dB
\end{eqnarray}
and
\begin{eqnarray}
\sigma^s_{ER}(E_{c.m.}) =&&\frac{\pi\hbar^2}{2\mu E_{c.m.}} \sum \limits ^{J_{max}}_{J=0} (2J+1)T(E_{c.m.},J) \nonumber  \\  && \times P_{CN}(E_{c.m.},J) W^s_{sur}(E_{c.m.},J),
\end{eqnarray}
respectively. The $\mu$ is the reduced mass of relative motion in colliding system. The transmission probability $T(E_{c.m.},J)$ is taken as zero and unity corresponding the incident energy $E_{c.m.}$ in the centre of mass frame below and above the summation value of attempting barrier $B$ and rotational energy at the relative angular momentum $J$. The $E_{1}$ and $J_{1}$ are the excitation energy and the angular momentum for the fragment (Z$_{1}$, N$_{1}$). The maximal angular momentum $J_{\max}$ is taken to be the grazing collision of two nuclei. The survival probability $W_{sur}$ of each fragment is evaluated with a statistical approach based on the Weisskopf evaporation theory \cite{Ch16}, in which the excited primary fragments are cooled in evaporation channels $s(Z_{s},N_{s})$ by $\gamma$-rays, light particles (neutrons, protons, $\alpha$ etc) in competition with the binary fission via $Z_{1}=Z^{'}_{1}-Z_{s}$ and $N_{1}=N^{'}_{1}-N_{s}$. The $P_{CN}(E_{c.m.},J)$ are fusion probability which sum over all the fragments probability located outside of BG (Businaro Gallone) point. The transferred cross section is smoothed with the barrier distribution and the function is taken as the Gaussian form of $f(B)=\frac{1}{N}\exp\left[-((B-B_{m})/\Delta)^{2}\right]$ with the normalization constant satisfying the unity relation $\int  f(B) dB=1$. The quantities $B_{m}$ and $\Delta$ are evaluated by $B_{m}=(B_{C}+B_{S})/2$ and $\Delta=(B_{C}-B_{S})/2$, respectively. The $B_{C}$ and $B_{S}$ are the Coulomb barrier at waist-to-waist orientation and the minimum barrier with varying the quadrupole deformation parameters of colliding partners.

\begin{figure}
\includegraphics[width=1.\linewidth]{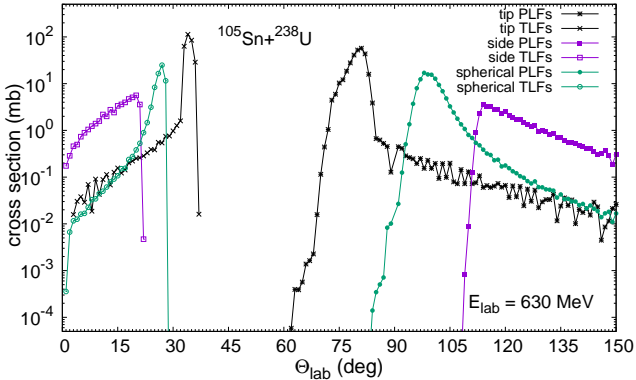}
\caption{\label{fig4} The angular distributions of the Sn-like and U-like products in the laboratory frame in the MNT reactions of $^{105}$Sn+$^{238}$U collisions at E$_{lab}$ = 6 MeV/nucleon with different collision orientations.}
\end{figure}

\begin{figure}
\includegraphics[width=0.99\linewidth]{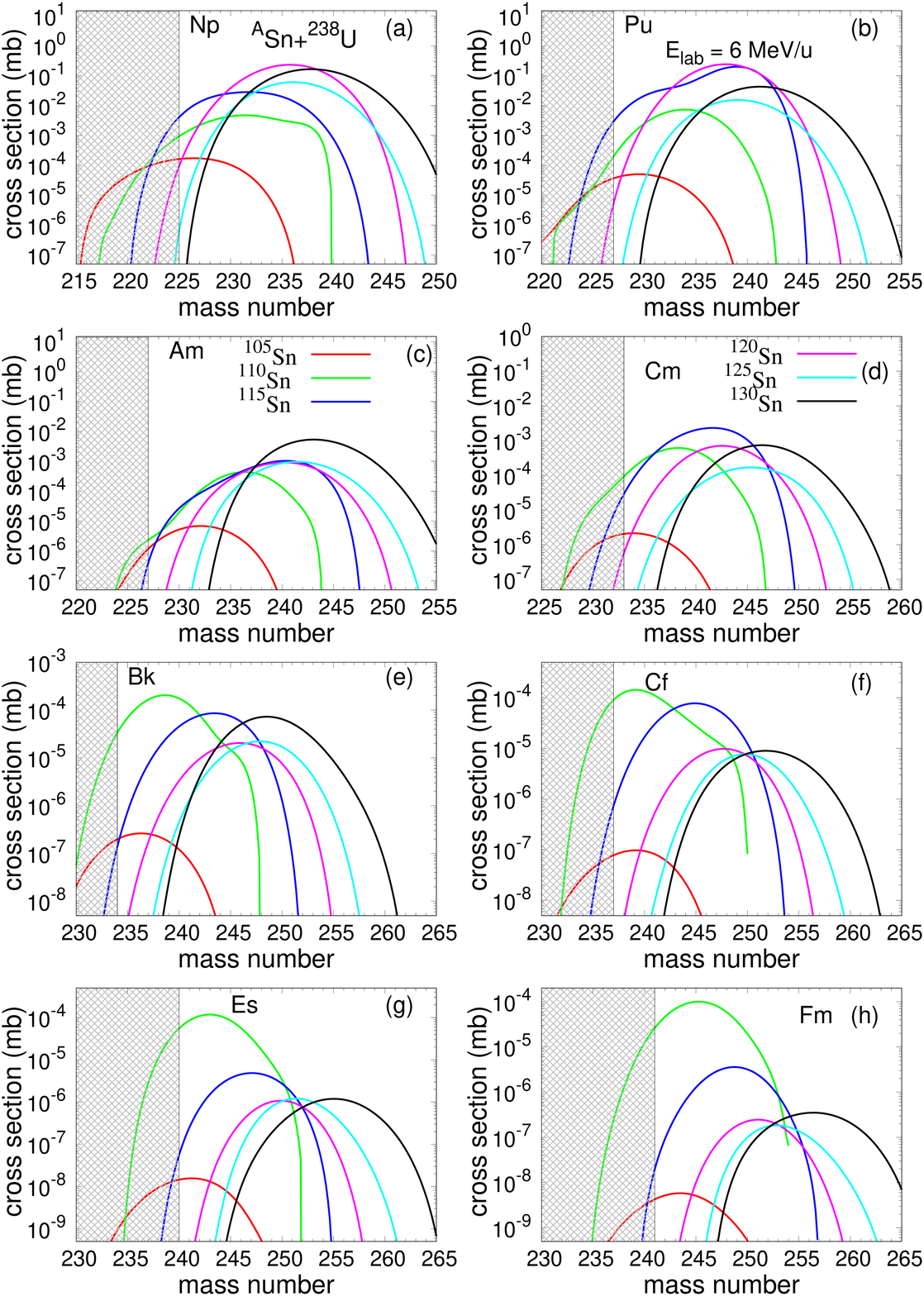}
\caption{\label{fig5} Cross sections for producing heavy neutron-deficient isotopes from the transfer reactions $^{105}$Sn + $^{238}$U (red line), $^{110}$Sn + $^{238}$U (green line), $^{115}$Sn + $^{238}$U (blue line), $^{120}$Sn + $^{238}$U (pink line), $^{125}$Sn + $^{238}$U (cyan line), $^{130}$Sn + $^{238}$U (black line) at the incident energy 6 MeV/nucleon. The grid line region indicates unknown isotopes.}
\end{figure}

\begin{figure}
\includegraphics[width=1\linewidth]{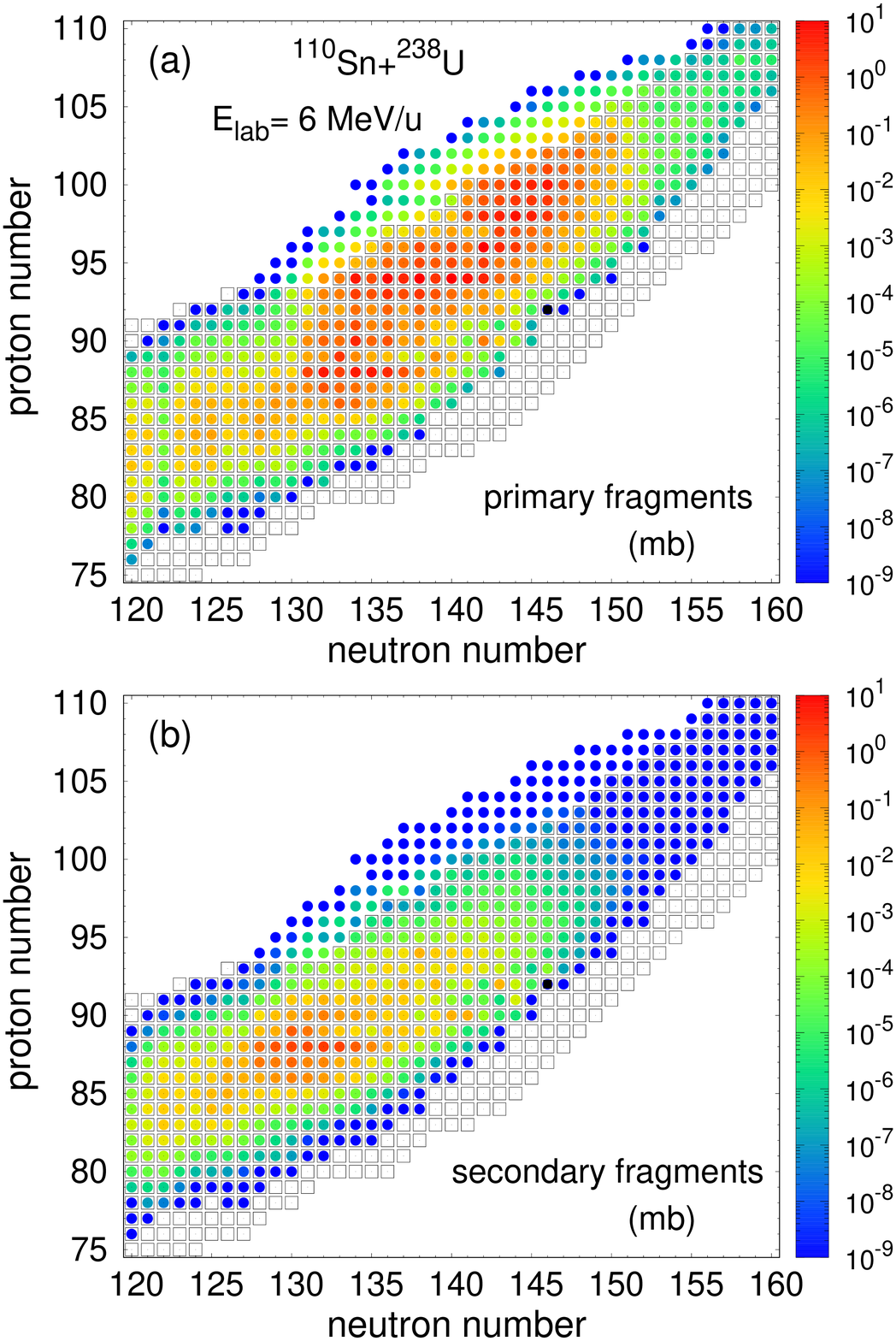}
\caption{\label{fig6} Contour plot of production cross sections as functions of neutron and proton numbers of primary fragments and secondary fragments in collisions of $^{105}$Sn + $^{238}$U at the incident energy $E_{lab}$ = 6 MeV/nucleon. The open squares and solid circles stand for known isotopes \cite{Wan17} and proton-rich unknown isotopes, respectively.}
\end{figure}

\begin{figure*}
\includegraphics[width=1\linewidth]{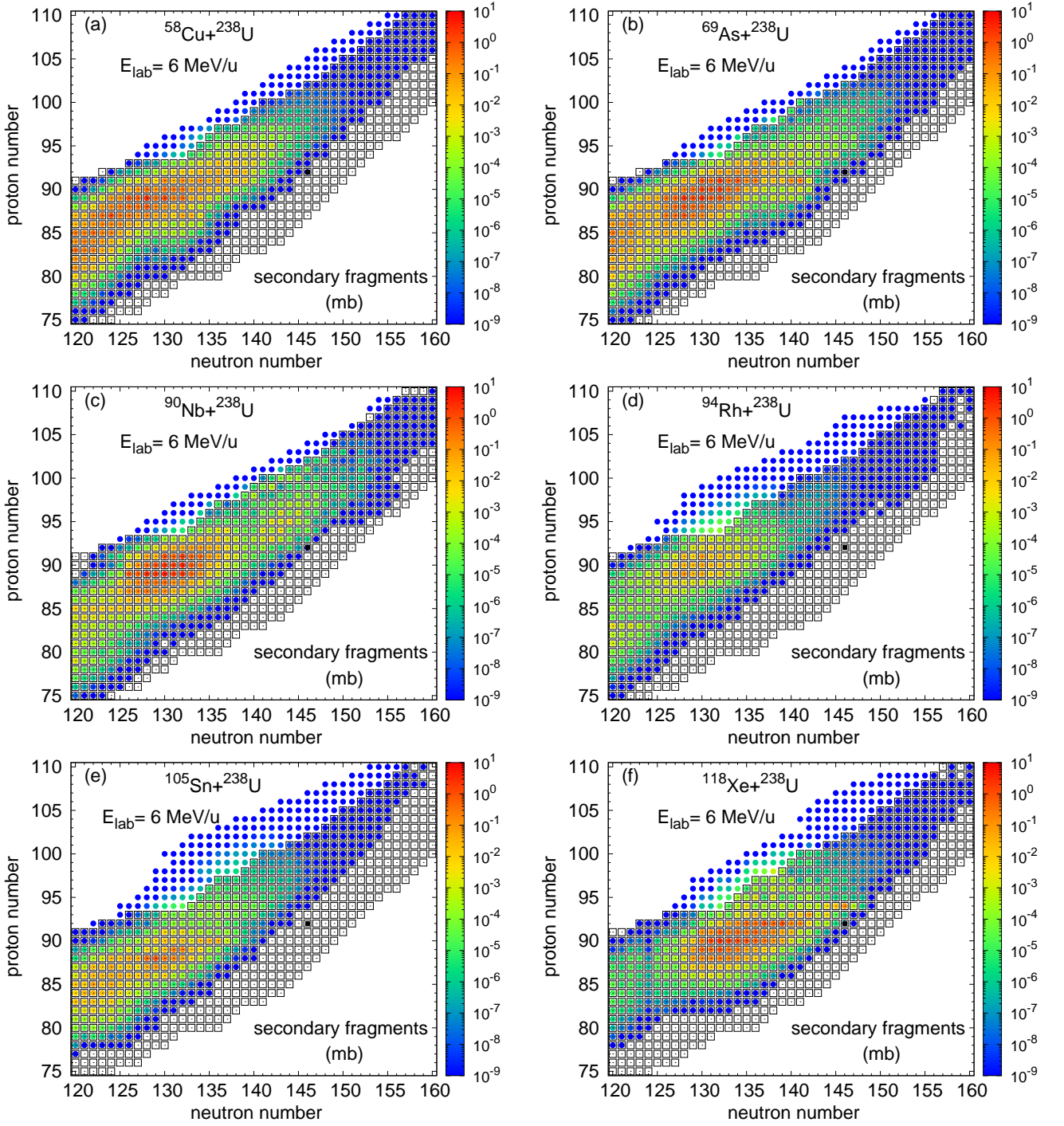}
\caption{\label{fig7} Production cross sections of final products in the MNT reactions with (a) $^{58}$Cu, (b) $^{69}$As, (c) $^{90}$Nb, (d) $^{94}$Rh, (e) $^{105}$Sn, and (f) $^{118}$Xe on $^{238}$U at the incident energy $E_{lab}$ = 6 MeV/nucleon. The entrance channels are marked by black solid squares.}
\end{figure*}

\begin{figure*}
\includegraphics[width=18 cm]{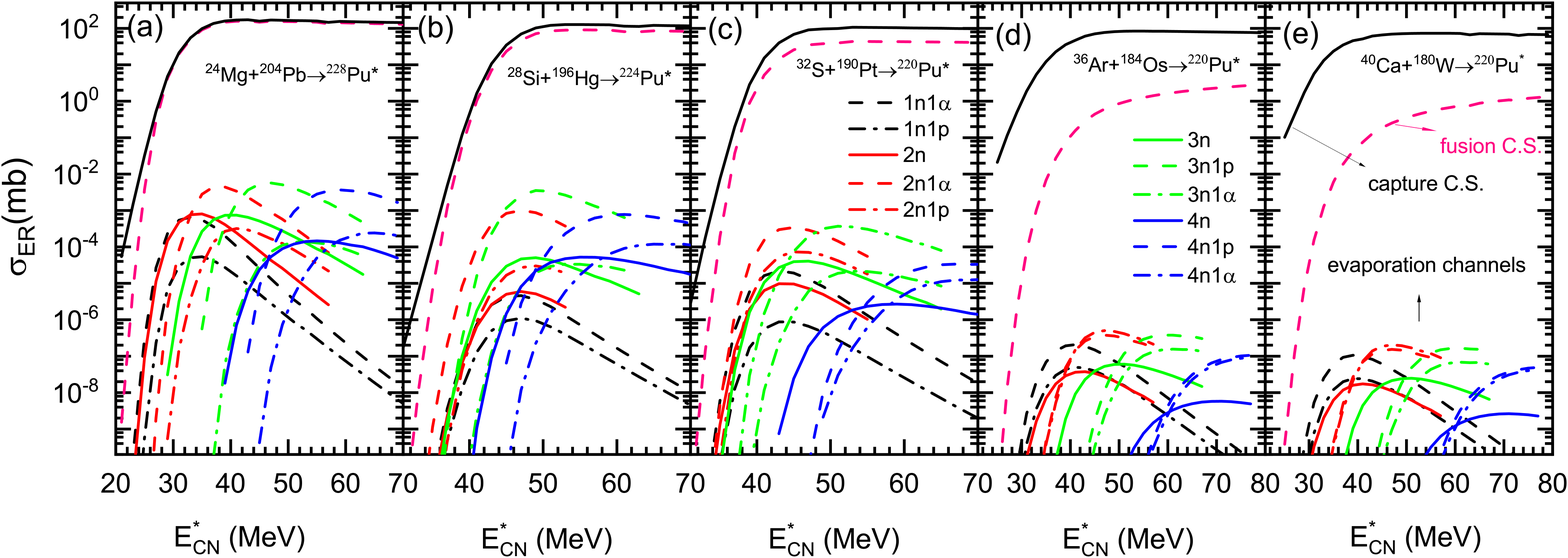}
\caption{\label{fig8} The fusion-evaporation reactions of (a) $^{24}$Mg + $^{204}$Pb, (b) $^{28}$Si + $^{196}$Hg, (c) $^{32}$S + $^{190}$Pt, (d) $^{36}$Ar + $^{184}$Os, and (e) $^{40}$Ca + $^{180}$W for producing the same compound nuclide Pu. The solid color lines, dashed lines and dot-dashed lines are the pure neutrons, neutrons mixed proton, neutrons mixed alpha channels, respectively.}
\end{figure*}

\begin{figure*}
\includegraphics[width=0.98\linewidth]{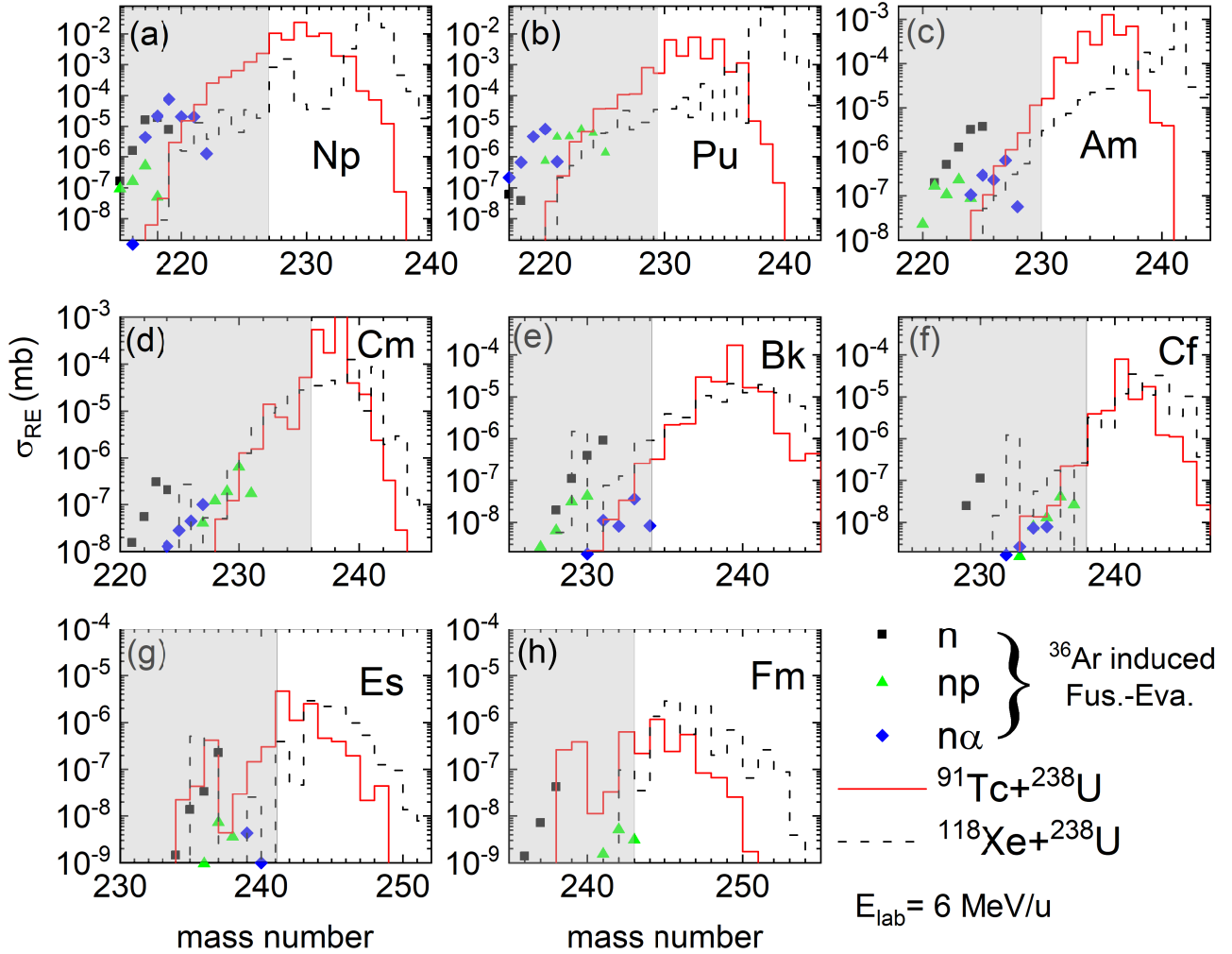}
\caption{\label{fig9} Comparison of the isotopic distributions of (a) Np, (b) Pu, (c) Am, (d) Cm, (e) Bk, (f) Cf, (g) Es and (h) Fm in the MNT reactions of $^{91}$Tc + $^{238}$U and $^{118}$Xe + $^{238}$U. The black, green and blue points stand for the pure neutron channels, neutron-proton mixing channels and alpha-neutron mixing channels in $^{36}$Ar induced fusion-evaporation reactions.}
\end{figure*}

\begin{table*}
\begin{ruledtabular}
\caption{Production cross sections of unknown neutron-deficient actinide isotopes with proton number Z and mass number A, predicted by the DNS model for the reactions of $^{58}$Cu + $^{238}$U, $^{69}$As + $^{238}$U, $^{90}$Nb + $^{238}$U, $^{94}$Rh + $^{238}$U,$^{105}$Sn + $^{238}$U, $^{118}$Xe + $^{238}$U around the Coulomb barrier energies. }
\label{tab1}
\begin{tabular}{c|c|c|c|c|c|c|c|c}
MNT & Np (mb) & Pu (mb) & Am (mb) &Cm (mb) & Bk (mb) & Cf (mb) & Es (mb) & Fm (mb) \\
\hline
 $^{58}$Cu+$^{238}$U & $^{225}$($2\times10^{-4}$) & $^{227}$($8\times10^{-6}$) & $^{229}$($1\times10^{-6}$) & $^{232}$($2\times10^{-5}$) & $^{233}$($8\times10^{-7}$) & $^{236}$($1\times10^{-8}$) & $^{240}$($2\times10^{-8}$) & $^{240}$($\textless10^{-9}$) \\

  & $^{224}$($1\times10^{-4}$) & $^{226}$($1\times10^{-6}$) & $^{228}$($8\times10^{-8}$) & $^{231}$($4\times10^{-6}$) & $^{232}$($8\times10^{-8}$) & $^{235}$($1\times10^{-9}$) & $^{239}$($5\times10^{-8}$) & $^{239}$($\textless10^{-9}$) \\

  & $^{223}$($5\times10^{-5}$) & $^{225}$($2\times10^{-7}$) & $^{227}$($2\times10^{-8}$) & $^{230}$($6\times10^{-7}$) & $^{231}$($1\times10^{-8}$) & $^{234}$($\textless10^{-9}$) & $^{238}$($4\times10^{-9}$) & $^{238}$($\textless10^{-9}$) \\

  & $^{222}$($2\times10^{-7}$) & $^{224}$($2\times10^{-8}$) & $^{226}$($1\times10^{-9}$) & $^{229}$($3\times10^{-8}$) & $^{230}$($\textless10^{-9}$) & $^{233}$($\textless10^{-9}$) & $^{237}$($\textless10^{-9}$) & $^{237}$($\textless10^{-9}$) \\

  & $^{221}$($1\times10^{-7}$) & $^{223}$($3\times10^{-9}$) & $^{225}$($\textless10^{-9}$) & $^{228}$($2\times10^{-9}$) & $^{229}$($\textless10^{-9}$) & $^{232}$($\textless10^{-9}$) & $^{236}$($\textless10^{-9}$) & $^{236}$($\textless10^{-9}$) \\

  & $^{220}$($6\times10^{-9}$) & $^{222}$($\textless10^{-9}$) & $^{224}$($\textless10^{-9}$) & $^{227}$($\textless10^{-9}$) & $^{228}$($\textless10^{-9}$) & $^{231}$($\textless10^{-9}$) & $^{235}$($\textless10^{-9}$) & $^{235}$($\textless10^{-9}$) \\
\hline
 $^{69}$As+$^{238}$U & $^{225}$($1\times10^{-3}$) & $^{227}$($2\times10^{-5}$) & $^{229}$($1\times10^{-5}$) & $^{232}$($4\times10^{-7}$) & $^{233}$($1\times10^{-6}$) & $^{236}$($5\times10^{-6}$) & $^{240}$($3\times10^{-8}$) & $^{240}$($1\times10^{-9}$) \\

   & $^{224}$($2\times10^{-4}$) & $^{226}$($7\times10^{-6}$) & $^{228}$($5\times10^{-7}$) & $^{231}$($2\times10^{-7}$) & $^{232}$($\textless10^{-9}$) & $^{235}$($6\times10^{-8}$) & $^{239}$($8\times10^{-6}$) & $^{239}$($\textless10^{-9}$) \\

  & $^{223}$($1\times10^{-4}$) & $^{225}$($1\times10^{-6}$) & $^{227}$($6\times10^{-7}$) & $^{230}$($2\times10^{-6}$) & $^{231}$($\textless10^{-9}$) & $^{234}$($\textless10^{-9}$) & $^{238}$($\textless10^{-9}$) & $^{238}$($\textless10^{-9}$) \\

  & $^{222}$($4\times10^{-7}$) & $^{224}$($6\times10^{-8}$) & $^{226}$($2\times10^{-9}$) & $^{229}$($1\times10^{-9}$) & $^{230}$($\textless10^{-9}$) & $^{233}$($\textless10^{-9}$) & $^{237}$($\textless10^{-9}$) & $^{237}$($\textless10^{-9}$) \\

  & $^{221}$($2\times10^{-7}$) & $^{223}$($6\times10^{-9}$) & $^{225}$($\textless10^{-9}$) & $^{228}$($\textless10^{-9}$) & $^{229}$($\textless10^{-9}$) & $^{232}$($\textless10^{-9}$) & $^{236}$($\textless10^{-9}$) & $^{236}$($\textless10^{-9}$) \\

  & $^{220}$($1\times10^{-8}$) & $^{222}$($\textless10^{-9}$) & $^{224}$($\textless10^{-9}$) & $^{227}$($\textless10^{-9}$) & $^{228}$($\textless10^{-9}$) & $^{231}$($\textless10^{-9}$) & $^{235}$($\textless10^{-9}$) & $^{235}$($\textless10^{-9}$) \\

  & $^{219}$($1\times10^{-9}$) & $^{221}$($\textless10^{-9}$) & $^{223}$($\textless10^{-9}$) & $^{226}$($\textless10^{-9}$) & $^{227}$($\textless10^{-9}$) & $^{230}$($\textless10^{-9}$) & $^{234}$($\textless10^{-9}$) & $^{234}$($\textless10^{-9}$) \\

  & $^{218}$($\textless10^{-9}$) & $^{220}$($\textless10^{-9}$) & $^{222}$($\textless10^{-9}$) & $^{225}$($\textless10^{-9}$) & $^{226}$($\textless10^{-9}$) & $^{229}$($\textless10^{-9}$) & $^{233}$($\textless10^{-9}$) & $^{233}$($\textless10^{-9}$) \\
 \hline
 $^{90}$Nb+$^{238}$U & $^{225}$($7\times10^{-3}$) & $^{227}$($2\times10^{-5}$) & $^{229}$($4\times10^{-6}$) & $^{232}$($2\times10^{-7}$) & $^{233}$($1\times10^{-5}$) & $^{236}$($1\times10^{-6}$) & $^{240}$($5\times10^{-5}$) & $^{240}$($\textless10^{-9}$) \\

  & $^{224}$($2\times10^{-4}$) & $^{226}$($5\times10^{-6}$) & $^{228}$($3\times10^{-7}$) & $^{231}$($7\times10^{-8}$) & $^{232}$($1\times10^{-7}$) & $^{235}$($\textless10^{-9}$) & $^{239}$($6\times10^{-6}$) & $^{239}$($\textless10^{-9}$) \\

  & $^{223}$($1\times10^{-4}$) & $^{225}$($2\times10^{-6}$) & $^{227}$($8\times10^{-7}$) & $^{230}$($9\times10^{-7}$) & $^{231}$($\textless10^{-9}$) & $^{234}$($\textless10^{-9}$) & $^{238}$($5\times10^{-8}$) & $^{238}$($\textless10^{-9}$) \\

  & $^{222}$($1\times10^{-6}$) & $^{224}$($8\times10^{-6}$) & $^{226}$($5\times10^{-9}$) & $^{229}$($1\times10^{-9}$) & $^{230}$($\textless10^{-9}$) & $^{233}$($\textless10^{-9}$) & $^{237}$($\textless10^{-9}$) & $^{237}$($\textless10^{-9}$) \\

  & $^{221}$($1\times10^{-6}$) & $^{223}$($1\times10^{-7}$) & $^{225}$($\textless10^{-9}$) & $^{228}$($\textless10^{-9}$) & $^{229}$($\textless10^{-9}$) & $^{232}$($\textless10^{-9}$) & $^{236}$($\textless10^{-9}$) & $^{236}$($\textless10^{-9}$) \\

  & $^{220}$($1\times10^{-8}$) & $^{222}$($2\times10^{-8}$) & $^{224}$($\textless10^{-9}$) & $^{227}$($\textless10^{-9}$) & $^{228}$($\textless10^{-9}$) & $^{231}$($\textless10^{-9}$) & $^{235}$($\textless10^{-9}$) & $^{235}$($\textless10^{-9}$) \\

  & $^{219}$($6\times10^{-9}$) & $^{221}$($\textless10^{-9}$) & $^{223}$($\textless10^{-9}$) & $^{226}$($\textless10^{-9}$) & $^{227}$($\textless10^{-9}$) & $^{230}$($\textless10^{-9}$) & $^{234}$($\textless10^{-9}$) & $^{234}$($\textless10^{-9}$) \\

  & $^{218}$($\textless10^{-9}$) & $^{220}$($\textless10^{-9}$) & $^{222}$($\textless10^{-9}$) & $^{225}$($\textless10^{-9}$) & $^{226}$($\textless10^{-9}$) & $^{229}$($\textless10^{-9}$) & $^{233}$($\textless10^{-9}$) & $^{233}$($\textless10^{-9}$) \\
  \hline

 $^{94}$Rh+$^{238}$U & $^{225}$($9\times10^{-4}$) & $^{227}$($7\times10^{-5}$) & $^{229}$($6\times10^{-5}$) & $^{232}$($5\times10^{-6}$) & $^{233}$($1\times10^{-6}$) & $^{236}$($3\times10^{-8}$) & $^{240}$($1\times10^{-8}$) & $^{240}$($\textless10^{-9}$) \\

 & $^{224}$($1\times10^{-3}$) & $^{226}$($1\times10^{-4}$) & $^{228}$($1\times10^{-5}$) & $^{231}$($2\times10^{-6}$) & $^{232}$($7\times10^{-7}$) & $^{235}$($1\times10^{-7}$) & $^{239}$($5\times10^{-9}$) & $^{239}$($\textless10^{-9}$) \\

  & $^{223}$($1\times10^{-3}$) & $^{225}$($7\times10^{-5}$) & $^{227}$($2\times10^{-5}$) & $^{230}$($6\times10^{-6}$) & $^{231}$($1\times10^{-6}$) & $^{234}$($2\times10^{-7}$) & $^{238}$($\textless10^{-9}$) & $^{238}$($\textless10^{-9}$) \\

  & $^{222}$($3\times10^{-5}$) & $^{224}$($2\times10^{-5}$) & $^{226}$($4\times10^{-6}$) & $^{229}$($1\times10^{-6}$) & $^{230}$($1\times10^{-7}$) & $^{233}$($4\times10^{-8}$) & $^{237}$($1\times10^{-8}$) & $^{237}$($\textless10^{-9}$) \\

  & $^{221}$($3\times10^{-5}$) & $^{223}$($8\times10^{-6}$) & $^{225}$($2\times10^{-6}$) & $^{228}$($1\times10^{-6}$) & $^{229}$($7\times10^{-8}$) & $^{232}$($1\times10^{-8}$) & $^{236}$($4\times10^{-9}$) & $^{236}$($\textless10^{-9}$) \\

  & $^{220}$($1\times10^{-6}$) & $^{222}$($2\times10^{-7}$) & $^{224}$($6\times10^{-8}$) & $^{227}$($1\times10^{-7}$) & $^{228}$($5\times10^{-9}$) & $^{231}$($2\times10^{-9}$) & $^{235}$($6\times10^{-9}$) & $^{235}$($\textless10^{-9}$) \\

  & $^{219}$($1\times10^{-8}$) & $^{221}$($1\times10^{-8}$) & $^{223}$($2\times10^{-8}$) & $^{226}$($6\times10^{-8}$) & $^{227}$($\textless10^{-9}$) & $^{230}$($\textless10^{-9}$) & $^{234}$($1\times10^{-9}$) & $^{234}$($\textless10^{-9}$) \\

  & $^{218}$($\textless10^{-9}$) & $^{220}$($\textless10^{-9}$) & $^{222}$($\textless10^{-9}$) & $^{225}$($\textless10^{-9}$) & $^{226}$($\textless10^{-9}$) & $^{229}$($\textless10^{-9}$) & $^{233}$($\textless10^{-9}$) & $^{233}$($\textless10^{-9}$) \\
  \hline
 $^{105}$Sn+$^{238}$U & $^{225}$($1\times10^{-4}$) & $^{227}$($9\times10^{-5}$) & $^{229}$($7\times10^{-6}$) & $^{232}$($2\times10^{-5}$) & $^{233}$($3\times10^{-6}$) & $^{236}$($2\times10^{-6}$) & $^{240}$($8\times10^{-8}$) & $^{240}$($1\times10^{-7}$) \\

   & $^{224}$($5\times10^{-5}$) & $^{226}$($2\times10^{-5}$) & $^{228}$($9\times10^{-7}$) & $^{231}$($6\times10^{-6}$) & $^{232}$($2\times10^{-8}$) & $^{235}$($6\times10^{-7}$) & $^{239}$($8\times10^{-7}$) & $^{239}$($1\times10^{-7}$) \\

  & $^{223}$($1\times10^{-4}$) & $^{225}$($1\times10^{-5}$) & $^{227}$($5\times10^{-7}$) & $^{230}$($3\times10^{-7}$) & $^{231}$($7\times10^{-9}$) & $^{234}$($9\times10^{-7}$) & $^{238}$($3\times10^{-7}$) & $^{238}$($1\times10^{-7}$) \\

  & $^{222}$($4\times10^{-7}$) & $^{224}$($3\times10^{-6}$) & $^{226}$($9\times10^{-8}$) & $^{229}$($6\times10^{-8}$) & $^{230}$($\textless10^{-9}$) & $^{233}$($1\times10^{-7}$) & $^{237}$($6\times10^{-7}$) & $^{237}$($3\times10^{-8}$) \\

  & $^{221}$($3\times10^{-7}$) & $^{223}$($3\times10^{-7}$) & $^{225}$($4\times10^{-8}$) & $^{228}$($1\times10^{-8}$) & $\textendash$ & $^{232}$($7\times10^{-8}$) & $^{236}$($1\times10^{-7}$) & $^{236}$($2\times10^{-8}$) \\

  & $^{220}$($\textless10^{-9}$) & $^{222}$($8\times10^{-9}$) & $^{224}$($\textless10^{-9}$) & $^{227}$($2\times10^{-9}$) & $\textendash$ & $^{231}$($\textless10^{-9}$) & $^{235}$($1\times10^{-7}$) & $^{235}$($2\times10^{-9}$) \\

  & $\textendash$ & $^{221}$($\textless10^{-9}$) & $\textendash$ & $^{226}$($\textless10^{-9}$) & $ \textendash $ & $ \textendash $ & $^{234}$($1\times10^{-8}$) & $^{234}$($\textless10^{-9}$) \\

  & $\textendash$ & $\textendash$  & $\textendash$ & $\textendash$ & $\textendash$ & $\textendash$ & $^{233}$($7\times10^{-9}$) & $\textendash$ \\

  & $\textendash$ & $\textendash$ & $\textendash$ & $\textendash$ & $\textendash$ & $\textendash$ & $^{232}$($\textless10^{-9}$) & $\textendash$ \\
  \hline

  $^{118}$Xe+$^{238}$U & $^{225}$($2\times10^{-4}$) & $^{227}$($2\times10^{-5}$) & $^{229}$($2\times10^{-4}$) & $^{232}$($2\times10^{-4}$) & $^{233}$($9\times10^{-5}$) & $^{236}$($9\times10^{-4}$) & $^{240}$($3\times10^{-6}$) & $^{240}$($2\times10^{-6}$) \\

      & $^{224}$($2\times10^{-5}$) & $^{226}$($3\times10^{-5}$) & $^{228}$($1\times10^{-5}$) & $^{231}$($6\times10^{-6}$) & $^{232}$($3\times10^{-6}$) & $^{235}$($8\times10^{-5}$) & $^{239}$($1\times10^{-5}$) & $^{239}$($5\times10^{-7}$) \\

  & $^{223}$($4\times10^{-5}$) & $^{225}$($5\times10^{-7}$) & $^{227}$($8\times10^{-6}$) & $^{230}$($1\times10^{-5}$) & $^{231}$($1\times10^{-6}$) & $^{234}$($5\times10^{-5}$) & $^{238}$($1\times10^{-6}$) & $^{238}$($3\times10^{-7}$) \\

  & $^{222}$($5\times10^{-8}$) & $^{224}$($4\times10^{-7}$) & $^{226}$($9\times10^{-8}$) & $^{229}$($4\times10^{-7}$) & $^{230}$($1\times10^{-8}$) & $^{233}$($2\times10^{-6}$) & $^{237}$($1\times10^{-6}$) & $^{237}$($\textless10^{-9}$) \\

  & $^{221}$($1\times10^{-8}$) & $^{223}$($4\times10^{-9}$) & $^{225}$($4\times10^{-8}$) & $^{228}$($8\times10^{-7}$) & $^{229}$($8\times10^{-8}$) & $^{232}$($\textless10^{-9}$) & $^{236}$($5\times10^{-7}$) & $^{236}$($\textendash$) \\

  & $^{220}$($\textless10^{-9}$) & $^{222}$($\textless10^{-9}$) & $^{224}$($\textless10^{-9}$) & $^{227}$($\textless10^{-9}$) & $^{228}$($\textless10^{-9}$) & $\textendash$ & $^{235}$($9\times10^{-7}$) & $^{235}$($\textendash$) \\

  & $\textendash$ & $\textendash$ & $\textendash$ & $\textendash$ & $\textendash$ & $\textendash$ & $1\times10^{-8}$ & $\textendash$ \\

  & $\textendash$ & $\textendash$ & $ \textendash$ & $\textendash$ & $\textendash$ & $\textendash$ & $\textless10^{-9}$ & $\textendash$ \\

  \end{tabular}
\end{ruledtabular}
\end{table*}

\begin{table*}
\begin{ruledtabular}
\caption{Cross sections of unknown proton-rich actinide isotopes with Z=93-100 predicted by the DNS model in the fusion-evaporation reactions for the $^{40}$Ca, $^{36}$Ar, $^{32}$S induced reactions with targets of $^{181}$Ta, $^{180}$W, $^{185}$Re, $^{184}$Os, $^{191}$Ir, $^{190}$Pt, $^{197}$Au, $^{196}$Hg, $^{203}$Tl, $^{204}$Pb, $^{209}$Bi. The evaporation channels are listed in the first column. The projectiles and targets are listed in the same rows.}
\label{tab2}
\begin{tabular}{c|c|c|c|c|c|c|c|c}
CFR & Np (mb) & Pu (mb) & Am (mb) &Cm (mb) & Bk (mb) & Cf (mb) & Es (mb) & Fm (mb) \\
\hline
  $^{40}$Ca+ &$^{181}$Ta & $^{180}$W &  $^{185}$Re  & $^{184}$Os & $^{191}$Ir & $^{190}$Pt & $^{197}$Au  &  $^{196}$Hg  \\

  2n & $^{219}  (1\times 10^{-6}) $ & $^{218} (1\times10^{-8})$ & $^{223} (6\times10^{-8})$ & $^{222} (4\times10^{-9})$ & $^{229} (5\times10^{-7})$ & $^{228} (7\times10^{-9})$ & $^{235} (1\times10^{-7})$ & $^{234} (2\times10^{-8})$\\

  3n & $^{218}  (3\times 10^{-6}) $ & $^{217} (2\times10^{-8})$ & $^{222} (6\times10^{-8})$ & $^{221} (2\times10^{-9})$ & $^{228} (7\times10^{-8})$ & $^{227} (2\times10^{-9})$ & $^{234} (1\times10^{-8})$ & $^{233} (1\times10^{-9})$\\

  4n & $^{217}  (4\times 10^{-6}) $ & $^{216} (2\times10^{-9})$ & $^{221} (8\times10^{-8})$ & $^{220} (\textless10^{-9})$ & $^{227} (1\times10^{-8})$ &                      $\textendash$   & $^{233} (4\times10^{-9})$ & $^{232} (\textless10^{-9})$\\

  5n & $^{216}  (5\times 10^{-7}) $ & $^{215} (\textless10^{-9})$ & $^{220} (1\times10^{-8})$ &                $\textendash$                 & $^{226} (2\times10^{-9})$ &                                                                      $\textendash$  & $^{232} (\textless10^{-9})$ & $\textendash$ \\

  6n & $^{215}  (6\times 10^{-8}) $ &         $  \textendash $                     & $^{219} (1\times10^{-9})$ &                 $\textendash$                & $^{225} (\textless10^{-6})$ &                                        $ \textendash$   &                $\textendash $               & $\textendash$ \\

  $^{40}$Ca+  & $^{180}$W &  $^{185}$Re  & $^{184}$Os & $^{191}$Ir & $^{190}$Pt & $^{197}$Au  &  $^{196}$Hg &$^{203}$Tl  \\

  1n1p & $^{218}  (2\times 10^{-8}) $ & $^{223} (5\times10^{-8})$ & $^{222} (1\times10^{-9})$ & $^{229} (1\times10^{-7})$ & $^{228}(8\times10^{-9})$ & $^{235} (2\times10^{-8})$ & $^{234} (6\times10^{-9})$ & $^{241} (1\times10^{-9})$\\

  2n1p & $^{217}  (1\times 10^{-7}) $ & $^{222} (1\times10^{-7})$ & $^{221} (6\times10^{-9})$ & $^{228} (1\times10^{-7})$ & $^{227} (\textless10^{-9})$ & $^{234} (2\times10^{-8})$ & $^{233} (2\times10^{-9})$ & $^{240} (1\times10^{-9})$\\

 3n1p & $^{216}  (2\times 10^{-7}) $ & $^{221} (2\times10^{-7})$ & $^{220} (2\times10^{-9})$ & $^{227} (5\times10^{-8})$ & $\textendash$ & $^{233} (5\times10^{-9})$ & $^{232} (\textless10^{-9})$ & $^{239} (\textless10^{-9})$\\

  4n1p & $^{215}  (6\times 10^{-8}) $ & $^{220} (3\times10^{-7})$ & $^{219} (2\times10^{-9})$ & $^{226} (2\times10^{-8})$ & $\textendash$ & $^{232} (\textless10^{-9})$ & $\textendash$ & $\textendash$\\

  5n1p & $^{214}  (4\times 10^{-8}) $ & $^{219} (2\times10^{-7})$ & $^{218} (\textless10^{-9})$ & $^{225} (8\times10^{-9})$ & $\textendash$ & $\textendash$ & $\textendash$ & $\textendash$\\

  $^{40}$Ca+  &  $^{185}$Re  & $^{184}$Os & $^{191}$Ir & $^{190}$Pt & $^{197}$Au  &  $^{196}$Hg &$^{203}$Tl & $^{204}$Pb  \\

  1n1$\alpha$ & $^{220}  (4\times 10^{-8}) $ & $^{219} (3\times10^{-8})$ & $^{226} (8\times10^{-8})$ & $^{225} (2\times10^{-8})$ & $^{232}(8\times10^{-9})$ & $^{231} (1\times10^{-8})$ & $^{239} (\textless10^{-9})$ & $^{239} (1\times10^{-9})$\\

  2n1$\alpha$ & $^{219} (2\times 10^{-6}) $ & $^{218} (1\times10^{-8})$ & $^{225} (1\times10^{-7})$ & $^{224} (1\times10^{-8})$ & $^{231} (2\times10^{-8})$ & $^{230} (1\times10^{-9})$ & $^{238} (1\times10^{-9})$ & $^{238} (\textless10^{-9})$\\

 3n1$\alpha$ & $^{218}  (1\times 10^{-6}) $ & $^{217} (1\times10^{-8})$ & $^{224} (5\times10^{-8})$ & $^{223} (1\times10^{-8})$ & $^{230} (3\times10^{-9})$ & $^{229} (\textless10^{-9})$ & $^{237} (\textless10^{-9})$ & $\textendash $\\

  4n1$\alpha$ & $^{217}  (1\times 10^{-6}) $ & $^{216} (3\times10^{-9})$ & $^{223} (2\times10^{-8})$ & $^{222} (6\times10^{-9})$ & $^{229} (\textless10^{-9})$ & $\textless$ & $\textendash $ & $\textendash $\\

  5n1$\alpha$ & $^{216}  (2\times 10^{-7}) $ & $^{215} (\textless10^{-9})$ & $^{222} (8\times10^{-9})$ & $^{221} (1\times10^{-9})$ &   $\textendash$
  & $\textendash$ & $ \textendash $ & $\textendash$\\

  5n1$\alpha$ & $^{215}  (6\times 10^{-9}) $ &                         $\textendash$.       & $^{221} (\textless10^{-9})$ & $^{220} (\textless10^{-9})$ & $\textendash$
  & $\textendash$ & $\textendash $ & $\textendash$\\
  \hline

  \hline
  $^{36}$Ar+ &$^{185}$Re  & $^{184}$Os & $^{191}$Ir & $^{190}$Pt & $^{197}$Au  &  $^{196}$Hg &$^{203}$Tl & $^{204}$Pb  \\

  2n & $^{219}  (7\times 10^{-6}) $ & $^{218} (3\times10^{-8})$ & $^{225} (3\times10^{-6})$ & $^{224} (2\times10^{-7})$ & $^{231} (9\times10^{-7})$ & $^{230} (1\times10^{-7})$ & $^{237} (2\times10^{-7})$ & $^{238} (4\times10^{-8})$\\

  3n & $^{218}  (1\times 10^{-5}) $ & $^{217} (6\times10^{-8})$ & $^{224} (3\times10^{-6})$ & $^{223} (3\times10^{-7})$ & $^{230} (4\times10^{-7})$ & $^{229} (2\times10^{-8})$ & $^{236} (3\times10^{-8})$ & $^{237} (7\times10^{-9})$\\

  4n & $^{217}  (1\times 10^{-5}) $ & $^{216} (5\times10^{-9})$ & $^{223} (1\times10^{-6})$ & $^{222} (5\times10^{-8})$ & $^{229} (1\times10^{-7})$ & $^{228} (1\times10^{-9})$ & $^{235} (1\times10^{-8})$ & $^{236} (1\times10^{-9})$\\

  5n & $^{216}  (1\times 10^{-6}) $ & $^{215} (\textless10^{-9})$ & $^{222} (5\times10^{-7})$ & $^{221} (1\times10^{-8})$ & $^{228} (1\times10^{-8})$ & $^{227} (\textless10^{-9})$ & $^{234} (1\times10^{-9})$ & $^{235} (\textless10^{-9})$\\

  6n & $^{215}  (1\times 10^{-7}) $ & $\textendash$                                & $^{221} (1\times10^{-7})$ & $^{220} (\textless10^{-9})$ & $^{227} (\textless10^{-9})$ & $\textendash$                                & $^{233} (\textless10^{-9})$ & $\textendash $\\

  $^{36}$Ar+   & $^{184}$Os & $^{191}$Ir & $^{190}$Pt & $^{197}$Au  &  $^{196}$Hg &$^{203}$Tl & $^{204}$Pb &$^{209}$Bi  \\

  1n1p & $^{218}  (5\times 10^{-8}) $ & $^{225} (1\times10^{-6})$ & $^{224} (8\times10^{-8})$ & $^{231} (1\times10^{-7})$ & $^{230}(4\times10^{-8})$ & $^{237} (2\times10^{-8})$ & $^{238} (3\times10^{-9})$ & $^{243} (3\times10^{-9})$\\

  2n1p & $^{217}  (5\times 10^{-7}) $ & $^{224} (5\times10^{-6})$ & $^{223} (2\times10^{-7})$ & $^{230} (6\times10^{-7})$ & $^{229} (3\times10^{-8})$ & $^{236} (4\times10^{-8})$ & $^{237} (7\times10^{-9})$ & $^{242} (5\times10^{-9})$\\

 3n1p & $^{216}  (1\times 10^{-7}) $ & $^{223} (7\times10^{-6})$ & $^{222} (1\times10^{-7})$ & $^{229} (1\times10^{-7})$ & $^{228} (6\times10^{-9})$ & $^{235} (1\times10^{-8})$ & $^{236} (\textless10^{-9})$ & $^{241} (1\times10^{-9})$\\

  4n1p & $^{215}  (9\times 10^{-8}) $ & $^{222} (4\times10^{-6})$ & $^{221} (1\times10^{-7})$ & $^{228} (1\times10^{-7})$ & $^{227} (2\times10^{-9})$ & $^{234} (7\times10^{-9})$ & $\textendash $                                & $^{240} (\textless 10^{-9})$\\

  5n1p & $^{214}  (5\times 10^{-9}) $ & $^{221} (4\times10^{-6})$ & $^{220} (2\times10^{-8})$ & $^{227} (4\times10^{-8})$ & $^{226} (\textless10^{-9})$ & $^{233} (1\times10^{-9})$ & $\textendash $                                 & $\textendash $                               \\

  $^{36}$Ar+  & $^{191}$Ir & $^{190}$Pt & $^{197}$Au  &  $^{196}$Hg &$^{203}$Tl & $^{204}$Pb &$^{209}$Bi &    \\

  1n1$\alpha$ & $^{222}  (1\times 10^{-6}) $ & $^{221} (7\times10^{-7})$ & $^{228} (5\times10^{-8})$ & $^{227} (1\times10^{-7})$ & $^{234}(8\times10^{-9})$ & $^{235} (7\times10^{-9})$ & $^{240} (1\times10^{-9})$ & \\

  2n1$\alpha$ & $^{221} (2\times 10^{-5}) $ & $^{220} (8\times10^{-6})$ & $^{227} (6\times10^{-7})$ & $^{226} (4\times10^{-8})$ & $^{233} (3\times10^{-8})$ & $^{234} (7\times10^{-9})$ & $^{239} (4\times10^{-9})$ & \\

 3n1$\alpha$ & $^{220}  (2\times 10^{-5}) $ & $^{219} (4\times10^{-6})$ & $^{226} (2\times10^{-7})$ & $^{225} (2\times10^{-8})$ & $^{232} (8\times10^{-9})$ & $^{233} (4\times10^{-6})$ & $^{238} (\textless10^{-9})$ & \\

  4n1$\alpha$ & $^{219}  (7\times 10^{-5}) $ & $^{218} (6\times10^{-7})$ & $^{225} (2\times10^{-7})$ & $^{224} (1\times10^{-8})$ & $^{231} (1\times10^{-8})$ & $^{232} (2\times10^{-9})$ & $\textendash $ & \\

  5n1$\alpha$ & $^{218}  (2\times 10^{-5}) $ & $^{217} (2\times10^{-7})$ & $^{224} (1\times10^{-7})$ & $^{223} (5\times10^{-9})$ & $^{230} (1\times10^{-9})$ & $^{231} (1\times10^{-9})$ & $\textendash $ &\\

  5n1$\alpha$ & $^{217}  (4\times 10^{-6}) $ & $^{216} (1\times10^{-7})$ & $^{223} (5\times10^{-9})$ & $^{222} (\textless10^{-9})$ & $^{229} (\textless10^{-9})$ & $^{230} (\textless10^{-9})$ & $\textendash $ & \\
  \hline

  $^{32}$S+& $^{191}$Ir & $^{190}$Pt & $^{197}$Au  &  $^{196}$Hg &$^{203}$Tl & $^{204}$Pb &$^{209}$Bi &   \\

  2n & $^{221}  (6\times 10^{-5}) $ & $^{220} (9\times10^{-6})$ & $^{227} (1\times10^{-5})$ & $^{226} (1\times10^{-6})$ & $^{233} (1\times10^{-5})$ & $^{234} (1\times10^{-6})$ & $^{239} (3\times10^{-7})$ &\\

  3n & $^{220}  (3\times 10^{-4}) $ & $^{219} (4\times10^{-5})$ & $^{226} (1\times10^{-5})$ & $^{225} (1\times10^{-6})$ & $^{232} (1\times10^{-6})$ & $^{233} (3\times10^{-7})$ & $^{238} (8\times10^{-8})$ & \\

  4n & $^{219}  (1\times 10^{-3}) $ & $^{218} (2\times10^{-6})$ & $^{225} (7\times10^{-6})$ & $^{224} (1\times10^{-7})$ & $^{231} (6\times10^{-7})$ & $^{232} (4\times10^{-8})$ & $^{237} (4\times10^{-8})$ &\\

  5n & $^{218}  (1\times 10^{-4}) $ & $^{217} (6\times10^{-7})$ & $^{224} (1\times10^{-6})$ & $^{223} (9\times10^{-8})$ & $^{230} (1\times10^{-7})$ & $^{231} (9\times10^{-9})$ & $^{236} (5\times10^{-9})$ & \\

  6n & $^{217}  (6\times 10^{-5}) $ & $^{216} (5\times10^{-9})$ & $^{223} (4\times10^{-7})$ & $^{222} (4\times10^{-9})$ & $^{229} (1\times10^{-8})$ & $^{230} (\textless10^{-9})$ & $^{235} (\textless10^{-9})$ & \\

  $^{32}$S+  & $^{190}$Pt & $^{197}$Au  &  $^{196}$Hg &$^{203}$Tl & $^{204}$Pb &$^{209}$Bi  \\

  1n1p & $^{220}  (8\times 10^{-7}) $ & $^{227} (2\times10^{-6})$ & $^{226} (5\times10^{-7})$ & $^{233} (1\times10^{-6})$ & $^{234}(1\times10^{-7})$ & $^{239} (2\times10^{-8})$  \\

  2n1p & $^{219}  (7\times 10^{-5}) $ & $^{226} (2\times10^{-5})$ & $^{225} (1\times10^{-6})$ & $^{232} (2\times10^{-6})$ & $^{233} (2\times10^{-7})$ & $^{238} (6\times10^{-8})$ \\

 3n1p & $^{218}  (2\times 10^{-5}) $ & $^{225} (1\times10^{-5})$ & $^{224} (3\times10^{-7})$ & $^{231} (8\times10^{-7})$ & $^{232} (3\times10^{-8})$ & $^{237} (2\times10^{-8})$  \\

  4n1p & $^{217}  (1\times 10^{-5}) $ & $^{224} (1\times10^{-5})$ & $^{223} (2\times10^{-7})$ & $^{230} (5\times10^{-7})$ & $^{231} (4\times10^{-8})$ & $^{236} (1\times10^{-8})$  \\

  5n1p & $^{216}  (1\times 10^{-5}) $ & $^{223} (1\times10^{-5})$ & $^{222} (1\times10^{-7})$ & $^{229} (1\times10^{-7})$ & $^{230} (5\times10^{-9})$ & $^{235} (5\times10^{-9})$  \\

  6n1p & $^{215}  (1\times 10^{-8}) $ & $^{222} (1\times10^{-6})$ & $^{221} (5\times10^{-9})$ & $^{228} (2\times10^{-9})$ & $^{229} (\textless10^{-9})$ & $^{234} (\textless10^{-9})$  \\

  $^{32}$S+  & $^{197}$Au  &  $^{196}$Hg &$^{203}$Tl & $^{204}$Pb &$^{209}$Bi  \\

  1n1$\alpha$ & $^{224}  (1\times 10^{-6}) $ & $^{223} (4\times10^{-6})$ & $^{230} (4\times10^{-7})$ & $^{231} (4\times10^{-7})$ & $^{236}(8\times10^{-9})$    \\

  2n1$\alpha$ & $^{223}  (6\times 10^{-5}) $ & $^{222} (1\times10^{-5})$ & $^{229} (2\times10^{-6})$ & $^{230} (4\times10^{-7})$ & $^{235} (5\times10^{-8})$    \\

 3n1$\alpha$ & $^{222}  (6\times 10^{-5}) $ & $^{221} (8\times10^{-6})$ & $^{228} (6\times10^{-7})$ & $^{229} (1\times10^{-7})$ & $^{234} (1\times10^{-8})$    \\

  4n1$\alpha$ & $^{221}  (7\times 10^{-5}) $ & $^{220} (8\times10^{-6})$ & $^{227} (1\times10^{-6})$ & $^{228} (8\times10^{-8})$ & $^{233} (1\times10^{-8})$    \\

  5n1$\alpha$ & $^{220}  (4\times 10^{-5}) $ & $^{219} (4\times10^{-6})$ & $^{226} (3\times10^{-7})$ & $^{227} (3\times10^{-8})$ & $^{232} (2\times10^{-8})$   \\

  5n1$\alpha$ & $^{219}  (6\times 10^{-5}) $ & $^{218} (1\times10^{-7})$ & $^{225} (3\times10^{-8})$ & $^{226} (\textless10^{-9})$ & $^{231} (6\times10^{-9})$  \\

  6n1$\alpha$ & $^{218}  (2\times 10^{-7}) $ & $^{217} (\textless10^{-9})$ & $^{224} (\textless10^{-9})$ & $\textendash $                            & $^{230} (\textless10^{-9})$  \\

  \end{tabular}
\end{ruledtabular}
\end{table*}
\begin{table*}

\begin{ruledtabular}
\caption{Same as Table \ref{sec2}, but for the $^{28}$Si, $^{24}$Mg induced reactions. }
\label{tab3}
\begin{tabular}{c|c|c|c|c|c|c|c|c}
CFR & Np (mb) & Pu (mb) & Am (mb) &Cm (mb) & Bk (mb) & Cf (mb) & Es (mb) & Fm (mb) \\
\hline

  $^{28}$Si+ & $^{197}$Au  &  $^{196}$Hg &  $^{203}$Tl & $^{204}$Pb & $^{209}$Bi   \\

  2n & $^{223}  (2\times 10^{-5}) $ & $^{222} (5\times10^{-6})$ & $^{229} (3\times10^{-5})$ & $^{230} (1\times10^{-5})$ & $^{235} (4\times10^{-6})$  \\

  3n & $^{222}  (3\times 10^{-4}) $ & $^{221} (5\times10^{-5})$ & $^{228} (5\times10^{-5})$ & $^{229} (7\times10^{-6})$ & $^{234} (4\times10^{-6})$  \\

  4n & $^{221}  (1\times 10^{-3}) $ & $^{220} (5\times10^{-5})$ & $^{227} (3\times10^{-5})$ & $^{228} (1\times10^{-6})$ & $^{233} (2\times10^{-6})$  \\

  5n & $^{220}  (3\times 10^{-4}) $ & $^{219} (1\times10^{-5})$ & $^{226} (7\times10^{-6})$ & $^{227} (5\times10^{-7})$ & $^{232} (3\times10^{-7})$  \\

  6n & $^{219}  (9\times 10^{-4}) $ & $^{218} (8\times10^{-7})$ & $^{225} (2\times10^{-6})$ & $^{226} (2\times10^{-8})$ & $^{231} (9\times10^{-8})$  \\

  7n & $^{218}  (2\times 10^{-5}) $ & $^{217} (\textless10^{-9})$ & $^{224} (1\times10^{-8})$ & $^{225} (\textless10^{-9})$ & $^{230} (\textless10^{-9})$  \\

  $^{28}$Si+   &  $^{196}$Hg &$^{203}$Tl & $^{204}$Pb &$^{209}$Bi  \\

  1n1p & $^{222}  (1\times 10^{-6}) $ & $^{229} (3\times10^{-6})$ & $^{230} (8\times10^{-7})$ & $^{235} (3\times10^{-7})$ \\

  2n1p & $^{221}  (3\times 10^{-5}) $ & $^{228} (4\times10^{-5})$ & $^{229} (6\times10^{-6})$ & $^{234} (2\times10^{-6})$  \\

 3n1p & $^{220}  (3\times 10^{-5}) $ & $^{227} (5\times10^{-5})$ & $^{228} (1\times10^{-6})$ & $^{233} (1\times10^{-6})$  \\

  4n1p & $^{219}  (1\times 10^{-4}) $ & $^{226} (3\times10^{-5})$ & $^{227} (2\times10^{-6})$ & $^{232} (1\times10^{-6})$  \\

  5n1p & $^{218}  (2\times 10^{-5}) $ & $^{225} (3\times10^{-5})$ & $^{226} (6\times10^{-7})$ & $^{231} (7\times10^{-7})$  \\

  6n1p & $^{217}  (3\times 10^{-6}) $ & $^{224} (5\times10^{-6})$ & $^{225} (3\times10^{-8})$ & $^{230} (3\times10^{-8})$  \\

  $^{28}$Si+    & $^{203}$Tl & $^{204}$Pb &$^{209}$Bi   \\

  1n1$\alpha$ & $^{226}  (2\times 10^{-6}) $ & $^{227} (3\times10^{-6})$ & $^{232} (1\times10^{-7})$  \\

  2n1$\alpha$ & $^{225}  (1\times 10^{-4}) $ & $^{226} (1\times10^{-5})$ & $^{231} (3\times10^{-6})$  \\

 3n1$\alpha$ & $^{224}  (1\times 10^{-4}) $ & $^{225} (1\times10^{-5})$ & $^{230} (1\times10^{-6})$  \\

  4n1$\alpha$ & $^{223}  (1\times 10^{-4}) $ & $^{224} (1\times10^{-5})$ & $^{229} (2\times10^{-6})$  \\

  5n1$\alpha$ & $^{222}  (1\times 10^{-4}) $ & $^{223} (1\times10^{-5})$ & $^{228} (1\times10^{-6})$  \\

  6n1$\alpha$ & $^{221}  (6\times 10^{-5}) $ & $^{222} (1\times10^{-6})$ & $^{227} (1\times10^{-7})$  \\

  7n1$\alpha$ & $^{220}  (1\times 10^{-6}) $ & $^{221} (6\times10^{-9})$ & $^{226} (\textless10^{-9})$  \\

    \hline

  $^{24}$Mg+ & $^{203}$Tl & $^{204}$Pb &$^{209}$Bi   \\

  2n & $^{225}  (2\times 10^{-3}) $ & $^{226} (8\times10^{-4})$ & $^{231} (9\times10^{-4})$ \\

  3n & $^{224}  (6\times 10^{-3}) $ & $^{225} (7\times10^{-4})$ & $^{230} (4\times10^{-4})$ \\

  4n & $^{223}  (6\times 10^{-3}) $ & $^{224} (1\times10^{-4})$ & $^{229} (2\times10^{-4})$ \\

  5n & $^{222}  (1\times 10^{-3}) $ & $^{223} (9\times10^{-5})$ & $^{228} (2\times10^{-5})$ \\

  6n & $^{221}  (1\times 10^{-3}) $ & $^{222} (2\times10^{-5})$ & $^{227} (1\times10^{-5})$ \\

  7n & $^{220}  (1\times 10^{-4}) $ & $^{221} (1\times10^{-6})$ & $^{226} (1\times10^{-7})$ \\

  $^{24}$Mg+   & $^{204}$Pb &$^{209}$Bi  \\

  1n1p & $^{226}  (8\times 10^{-7}) $ & $^{231} (5\times10^{-5})$ \\

  2n1p & $^{225}  (3\times 10^{-4}) $ & $^{230} (2\times10^{-4})$ \\

  3n1p & $^{224}  (1\times 10^{-4}) $ & $^{229} (4\times10^{-4})$ \\

  4n1p & $^{223}  (2\times 10^{-4}) $ & $^{228} (9\times10^{-5})$ \\

  5n1p & $^{222}  (1\times 10^{-4}) $ & $^{227} (1\times10^{-4})$ \\

  6n1p & $^{221}  (5\times 10^{-5}) $ & $^{226} (1\times10^{-5})$ \\

  7n1p & $^{220}  (3\times 10^{-7}) $ & $^{225} (1\times10^{-7})$ \\

  $^{24}$Mg+  &$^{209}$Bi   \\

  1n1$\alpha$ & $^{228}  (2\times 10^{-5}) $ \\

  2n1$\alpha$ & $^{227}  (5\times 10^{-4}) $ \\

  3n1$\alpha$ & $^{226}  (2\times 10^{-4}) $ \\

  4n1$\alpha$ & $^{225}  (3\times 10^{-4}) $ \\

  5n1$\alpha$ & $^{224}  (2\times 10^{-4}) $  \\

  6n1$\alpha$ & $^{223}  (2\times 10^{-4}) $ \\

  7n1$\alpha$ & $^{222}  (5\times 10^{-6}) $ \\
  \end{tabular}
\end{ruledtabular}
\end{table*}

\section{Results and discussion}\label{sec3}
The complete fusion reaction mechanism has been used to synthesize many new heavy and superheavy nuclei in experimentally. Recently, a renew interest, the damped collisions of two heavy nuclei were investigated and motivated for producing heavy isotopes, in particular for new nuclides close to proton and neutron-rich drip lines. The DNS model can nicely reproduce the production cross sections of fusion-evaporation products and MNT yields \cite{Ba19,Ba18,Gu17,Ch18,Kr13,Ch20,Fe17}. The fragment yields in the MNT reactions are related to the emission angle in the laboratory system. It was observed that the clusters formed in the massive transfer reactions were emitted anisotropically \cite{Ji80}. A prediction of the polar angle structure for the MNT fragments is helpful for managing the detector system in experiments. The emission angle of the reaction products is helpful for arranging detectors in experiments. We use a deflection function method to evaluate the fragment angle which is related to the mass of fragment, angular momentum and incident energy. The deflection angle is composed of the Coulomb and nuclear interaction \cite{Wo78,Ch20}. Shown in figure \ref{fig4} is the PLF and TLF angular distributions of primary fragments within transferring 20 nucleons in the reaction of $^{105}$Sn + $^{238}$U at the laboratory incident energy of $E_{lab}$ = 6 MeV/nucleon. The emission of MNT fragments is associated with the collision orientation, i.e., the peak varying from the angle of 80$^{o}$ to 110$^{o}$ with the tip-tip to side-side orientation for the PLFs. The PLFs are distributed in the large polar angles in comparison with the TLFs owing to the contribution of low angular momenta.

Shown in Fig. \ref{fig5} is the cross sections for isotopes Z=93-100 in the MNT reactions of tin isotopes induced $^{238}$U collisions at the laboratory energy of $E_{lab} $= 6 MeV/nucleon. The projectile nuclei are $^{105}$Sn, $^{110}$Sn,$^{115}Sn$, $^{120}$Sn, $^{125}$Sn, $^{130}$Sn. It is interesting to compare the production cross sections for different projectiles Sn bombarded the same target $^{238}$U through the MNT reaction. For the colliding systems $^{105, 110, 115, 120, 125,130}$Sn + $^{238}$U, the neptunium (Np), plutonium (Pu), americium (Am), curium (Cm), berkelium (Bk), californium (Cf), einsteinium (Es), fermium (Fm) neutron deficient isotopes may be created by transferring one to eight protons from projectile to target nuclei and a few neutron transfer in inverse process. The calculated production cross sections of neutron-deficient isotopes Z=93-100 increase with decreasing the N/Z ratios of Sn isotopes. The more neutron-poor isotopes are favorable for the new isotope formation in the MNT reactions. The grid region indicates unknown neutron-deficient isotopes as shown in FIG. \ref{fig5}. The reaction system with smaller N/Z ratio enhances the formation of proton-rich actinide nuclides. For example, the reactions induced by $^{115}$Sn are favorable for producing unknown neutron-deficient $^{227}$Np and $^{233}$Pu with the cross sections of 10 $\mu$ b and 96 $\mu$ b, respectively. While the bombardment of $^{110}$Sn on $^{238}$U lead to the production of $^{234}$Am, $^{237}$Cm, $^{234}$Bk, $^{237}$Cf, $^{240}$Es, and $^{241}$Fm with the cross sections of 138 nb, 350 nb, 149 pb, 717 pb, 619 pb, and 167 pb, respectively. The difference of $^{110}$Sn and $^{115}$Sn induced reactions is caused from the deformation effect.

Neutron-deficient Sn isotopes can be generated by the proton or neutron induced asymmetric fission of actinide nuclide, for instance, the radioactive beam facilities, Beijing Rare Ion Beam Facility (BRIF) and the future Beijing Isotope-Separation on Line (BISOL). The contour plot of primary and secondary fragments (Z$>$ 75, N $>$ 120) in collision of $^{110}$Sn + $^{238}$U at $E_{lab}$ = 6 MeV/nucleon are calculated as shown in FIG. \ref{fig6}. The open squares and solid circles stand for known isotopes with the mass table \cite{Wan17} and proton-rich unknown isotopes, respectively. The primary fragments are produced on the neutron-deficient side caused by isospin relaxation. The de-excitation process moves the fragments to the $\beta$-stability line and even the neutron-rich side through emitting charge particles. The solid color circles outside open squares are predicted unknown neutron-deficient isotopes. It is obvious that the de-excitation process reduces the mass region and a number of proton-rich nuclides might be created via the MNT reactions.

It is of interest to compare the production cross section from different projectile isotopes bombarded the same target through the MNT reactions. The proton-rich nuclides $^{58}$Cu, $^{69}$As, $^{90}$Nb, $^{94}$Rh, $^{105}$Sn and $^{118}$Xe are chosen, which might be available for the neutron-deficient radioactive beams generated in the radioactive beam facilities. FIG. \ref{fig7} shows the production cross section of final fragments in collisions of $^{58}$Cu, $^{69}$As, $^{90}$Nb, $^{94}$Rh, $^{105}$Sn, $^{118}$Xe on $^{238}$U at incident energy $E_{lab}$ = 6 MeV/nucleon. It is obvious that the isotopic distribution width increases with the projectile mass. The solid color circle without open square are the predicted new neutron-deficient isotopes that listed in Table \ref{tab1}. The unknown neutron-deficient isotopes proton number increases with increasing projectile mass. The calculation of $^{118}$Xe induced reaction in favor to producing neutron-deficient isotopes Z=98-100. The $^{94}$Rh induced reaction is advantageous in producing neutron-deficient isotopes of Z=93-97.

Figure \ref{fig8} depicts that calculated evaporation residual cross section for producing neutron-deficient compound nucleus Pu from different projectile-target combinations through fusion-evaporation reactions. The black solid lines and pink dashed lines are the capture cross sections and fusion cross sections, respectively. One can see that the capture cross sections of the four systems are almost the same, because their Coulomb barriers are changing slightly. Their fusion cross sections are dropping rapidly with decreasing mass asymmetry caused by higher inner barrier. In the figure, the black dashed line and black dash-dotted line are the 1n1$\alpha$, 1n1p channels, respectively. The red solid line, red dashed line and red dash-dotted line indicate 2n, 2n1$\alpha$, 2n1p channels, respectively. The green solid line, green dashed line, green dash- dotted line are the 3n, 3n1$\alpha$, 3n1p channels, respectively. The blue solid line, blue dashed line and blue dash-dotted line stand for 4n, 4n1$\alpha$, 4n1p channels, respectively. The combined channels with the charged particles are of significance in the decay process and are the main way for proton-rich nuclide production.

The calculated production cross sections of neutron-deficient actinide nuclei with Z=93-100 through fusion-evaporation and multinucleon transfer reaction are in FIG. \ref{fig9}. The grey region indicates unknown neutron-deficient actinide isotopes. The black solid line and red solid line are $^{118}$Xe + $^{238}$U and $^{91}$Tc + $^{238}$U reactions, respectively. The reaction $^{91}$Tc + $^{238}$U takes an advantage to produce unknown neutron-deficient nuclei with Z=93-94, in comparsion of reaction $^{118}$Xe + $^{238}$U, that is a favor to produce unknown neutron-deficient nuclei with Z=95-100. The black solid square, green solid triangle and blue solid square stand for pure neutron channels, neutron mixed proton channels and neutron mixed alpha channels from $^{36}$Ar induced fusion-evaporation reactions. From $^{36}$Ar induced fusion-evaporation reactions, we found that synthesis of unknown neutron-deficient nuclei with Z=93-94 prefer neutron mixed alpha channels, while pure neutron channels are favor to producing unknown neutron-deficient nuclei with Z=95-100. Through comparing the production cross section via the fusion-evaporation and multinucleon transfer reactions, we found that fusion-evaporation reactions are still a promising way to produce neutron-deficient actinide nuclei. Moreover, the MNT reactions are favorable for creating the proton-rich isotopes within the large mass region.

The production cross section of new neutron-deficient nuclei with Z=93-100 are estimated via the fusion-evaporation reactions as shown in Table~\ref{tab2} and Table \ref{tab3} for the systems of $^{40}$Ca, $^{36}$Ar, $^{32}$S, $^{28}$Si, $^{24}$Mg induced fusion reactions. For producing neutron-deficient actinide nuclei, charge evaporation channels play an important role in de-excitation process. The products from charge evaporation channels are not very neutron-deficient, in comparision with pure neutrons channels. The MNT reactions with neutron-deficient radioactive beams may also produce new neutron-deficient isotopes. The production cross section of new neutron-deficient nuclei from MNT reaction are equivalent to that from fusion-evaporation reaction as shown in Table \ref{tab1}. Their production cross section at the level of pb to mb is feasible for measurements in laboratories. New neutron-deficient nuclei produced through MNT reactions are broader, compared with the fusion-evaporation reactions. Further measurements are expected in the future experient.

\section{Conclusions}\label{sec4}
In summary, the production of neutron-deficient actinide isotopes with the charge number of Z=93-100 has been thoroughly investigated within the DNS model through fusion-evaporation and multinucleon transfer reactions. For the MNT reactions, the systems of $^{105, 110,115,120,125,130}$Sn $^{58}$Cu, $^{69}$As, $^{90}$Nb, $^{94}$Rh and $^{118}$Xe bombarding $^{238}$U around Coulomb barrier energies are chosen. The $^{40}$Ca, $^{36}$Ar, $^{32}$S, $^{28}$Si and $^{24}$Mg induced fusion reactions are selected for comparison. The valley shape of the PES influences the formation of primary fragments and leads to the production of neutron-deficient isotopes. The de-excitation process shifts the proton excess of fragments towards the $\beta$-stability line. The isospin relaxation in the nucleon transfer is coupled to the dissipation of relative energy and angular momentum of colliding system. The fragment yields are associated with nuclear shapes of the colliding nuclei and details of the potential energy surface in the MNT reactions.

Production of proton-rich actinide isotopes are relying strongly on the projectile-target mass asymmetry in the FE reactions. The charge particles evaporation channels play an important role on final production cross section. The anisotropy emission of MNT fragments is associated with the incident energy and deformation of colliding system. The angular distribution of the PLFs is shifted to the forward region with increasing the Coulomb barrier. However, that of TLFs exhibits an opposite trend. The total kinetic energies and angular spectra of primary fragments are highly dependent on colliding orientations. The distribution width for transferring neutrons is broader in the tip-tip collision for the deformed reaction system.

Production cross sections are highly dependent on projectile isotopes in the MNT reactions. The new proton-rich actinides are related to the N/Z ratio of reaction system. The neutron-deficient nuclides $^{110}$Sn and $^{118}$Xe induced reactions are favorable for producing heavy neutron-deficient isotopes with the elements of Z=95-100. Furthermore, the $^{94}$Rh induced reaction  $^{94}$Rh+$^{238}$U is better for producing new neutron-deficient Np, Pu. The numerous unknown neutron-deficient nuclei from Z=93 to Z=100 are predicted with the production cross sections via the MNT and FE reactions, which are listed in Table \ref{tab1}, \ref{tab2} and \ref{tab3}. The FE reactions are still most promising to synthesize new neutron-deficient actinide nuclei. In addition, the MNT reactions with radioactive beams provide an alternative way, which has the advantage of a wide region of new isotopes.

\section{Acknowledgements}
This work was supported by the National Natural Science Foundation of China (Projects No. 11722546 and No. 11675226) and the Talent Program of South China University of Technology.

\end{document}